\documentclass[
 reprint,
superscriptaddress,
 amsmath,amssymb,
 aps,pre, 
]{revtex4-1}

\usepackage{color}
\usepackage{mathtools}
\usepackage{lipsum}
\usepackage{hyperref}
\usepackage{amsmath,amssymb}
\usepackage{graphicx}
\usepackage{dcolumn}
\usepackage{bm}
\usepackage{float}
\usepackage{epstopdf}
\usepackage{textcomp}
\usepackage[shortlabels]{enumitem}
\usepackage{verbatim}
\usepackage{xfrac}

\def\be{\begin{equation}}
\def\ee{\end{equation}}
\def\bea{\begin{eqnarray}}
\def\eea{\end{eqnarray}}
\def\bsplit{\begin{split}}
\def\esplit{\end{split}}

\def\p{\partial} 
\def\nn{\nonumber}
\def\f{\frac}

\def\fc{\vartheta} 

\def\l{\left(}
\def\r{\right)}
\def\la{\langle}
\def\ra{\rangle}

\def\ddx{\partial_x}

\def\ddt{\partial_t}
\def\dx{{\dot x}}

\def\dx{{\dot x}}
\def\m{\Delta\mu}

\newcommand{\taumass}{\hat{\tau}_m}
\newcommand{\refn}[1]{Eq. (\ref{#1})}
\newcommand{\pa}{\partial}

\newcommand{\ncbs}{\affiliation{Simons Centre for the Study of Living Machines, National Centre for Biological Sciences,TIFR, Bangalore 560065, India}}
\newcommand{\mbi}{\affiliation{Mechanobiology Institute and Department of Biological Sciences, National University of Singapore, 117411 Singapore}}
\newcommand{\curie}{\affiliation{Laboratoire Physico Chimie Curie, Institut Curie, PSL Research University, CNRS UMR168, 75005 Paris, France.}}

\begin{document}

\graphicspath{{figures/}}


\title{Soft inclusion in a confined fluctuating active gel}

\author{Amit Singh Vishen}
\ncbs
\author{J. F. Rupprecht} 
\mbi
\author{G.V. Shivashankar}
\mbi
\author{Jacques Prost}
\mbi
\curie
\author{Madan Rao}
\ncbs

\date{\today}
           
\begin{abstract}
We study stochastic dynamics of a point and extended inclusion within a one dimensional confined active viscoelastic gel. We show that the dynamics of a point inclusion can be described by  a Langevin equation 
with a confining potential and multiplicative noise. Using a systematic adiabatic elimination over the fast variables, we arrive at an overdamped equation with a proper definition of the multiplicative noise.
To highlight various features and to appeal to different biological contexts, we treat the inclusion in turn as a rigid extended element, an elastic element and a viscoelastic (Kelvin-Voigt) element.   
The dynamics for the shape and position of the extended inclusion can be described by coupled Langevin equations. Deriving exact expressions for the corresponding steady state probability distributions, we find that the active noise induces an attraction to the edges of the confining domain. In the presence of a competing centering force, we find that the shape of the probability distribution exhibits a sharp transition upon varying the amplitude of the active noise.
Our results could help understanding the positioning and deformability of biological inclusions, eg. organelles in cells, or nucleus and cells within tissues.

\end{abstract}

\maketitle

\section{Introduction}
The collective fluctuating dynamics of particles in a medium, each of which are driven out of equilibrium by dissipation of energy, 
is a fundamentally new branch of statistical physics \cite{Activereview}, with deep implications for the physics of cells and tissues \cite{Activereview,Prost2015}.
More recent studies have focussed on the effects of confinement on active matter, especially on the nature of forces on fixed or deformable boundaries,
as a result of specific boundary conditions \cite{Solon2015}.

However, the role of fluctuations in generating novel forces at the boundaries of confined active suspensions, have not
received adequate attention. Such studies have potential implications for the fluctuating dynamics of organelles embedded within the cell cytoplasm, described as an
active actomyosin suspension.

In this paper, we study the positioning and shape dynamics of a deformable inclusion embedded in a confined active gel. 
Formulated in this general way, our study is applicable to a variety of in-vivo and in-vitro contexts - (i) the positioning and shape fluctuations of the nucleus (or other localized organelles) within a cell \cite{Dupin2011, Gundersen2013, Morris2002, Makhija2016, Talwar2013}, (ii) the dynamics of large colloidal particles embedded in an active medium \cite{Weihs2006, Hameed2012,Lau2003,Fodor2015}, (iii) the positioning and dynamics of nuclei in multi-nucleated cells \cite{Mazumdar2002},
 (iv) the positioning and segregation of chromosomes within a nucleus \cite{Cremer2001, Bruinsma2014, Weber2012, Ganai2014} and  (v) the fluctuations of a cell / cell-junction within a 
developing tissue \cite{Barton2017, Curran2017}.

For concreteness, we consider that the inclusion represents the cell nucleus that is connected to the cell cortex, whose boundary is held fixed, for instance by 
surface attachment on a micro-patterned substrate coated with fibronectin \cite{Makhija2016, Li2014, Rupprecht2017} (Fig.\,\ref{fig:schematic}(a)).

At a conceptual level and to make general comparisons with experiments such as \cite{Makhija2016}, it suffices to study the problem in one dimension (1d). 
We start by setting up equations for a point inclusion in a viscoelastic medium in Sect.\,\ref{sec:noise_limit}, and adiabatically eliminate the fast variable to obtain an overdamped Langevin equation. We then setup the general equations valid for any type of inclusion in Sect.\,\ref{sec:basicequation}.
To highlight different aspects we systematically treat the inclusion as being a rigid element (Sect.\,\ref{sec:rigid}), an elastic element (Sect.\,\ref{sec:elastic}) and a viscoelastic (Kelvin-Voigt) element 
(Sect.\,\ref{sec:viscoelastic}).

We find that in general, the active dynamics of the position and shape of the inclusion are described by coupled Langevin equations with a confining potential and {\it multiplicative} noise.
Solving the Fokker-Planck equations corresponding to the Langevin description,  we obtain the exact steady state distributions for the position and size of the inclusion.

Our analysis shows that there are {\it sharp  transitions} in the nature of the probability distribution, as a function of the relative strength of the confining potential and active noise. 
The cell boundary influences 
the steady state position, and size of the inclusion.
The position, and size fluctuations of the inclusion are also driven by these active stress
fluctuations in the active fluid. 

This paper follows the Letter \cite{Rupprecht2017} which contains a comparison to experiments on the positioning and fluctuations of the nucleus in cells confined in micro-patterned surfaces.

The dynamics of the inclusion embedded in the fluctuating active fluid can be viewed as an active Casimir effect \cite{Bartolo2003, Ray2014, Parra-Rojas2014}.
 We show that this fluctuation-induced attraction depends on both the intensity of the active noise and on the hydrodynamic interactions of the inclusion with the boundaries. 
We find that the effect of such Casimir-type forces on the boundary confining a dilute active suspension, extends over large scales (Sect.\,\ref{sec:force}).

\section{Point inclusion in active Viscoelastic gel}\label{sec:noise_limit}

We consider a point inclusion  of mass $m$ embedded in a compressible 1d active gel confined between hard walls at $x=\pm L$. We denote the viscous fluid regions to the left and right of the inclusion as $(L)$ and $(R)$, respectively.

The hydrodynamic variables describing the bulk are the actomyosin concentration $c$ and the hydrodynamic velocity $v$. 
The velocity field of this active viscoelastic gel is determined by local force balance, $\ddx \sigma(x,t) = 0$. We express the local stress as
\be
(1 + \tau_v \partial_t)\sigma^{L,R} = \eta_c \ddx v - \zeta \m c + \fc_A + \fc_T,
\label{eq:stress1_Maxwell}
\ee
where $\tau_v$ is the maxwell relaxation time,  $\eta_c$ is a one-dimensional cortical viscosity, $\zeta \m < 0$ is the active contractile stress \cite{Activereview, Kruse2005, Prost2015}.
Here, $\fc$ represent stress fluctuations which may either be of thermal or active origin. 
Under the assumption of a constant viscosity $\eta_c$, the fluctuation-dissipation relation imposes thermal fluctuations should be delta-correlated: 
$\la \fc_T(x,t) \fc_T(x', t')\ra =  2\Lambda_T \delta(x - x') \delta(t-t')$, where $\Lambda_T = k_b T \eta_c$,  with $T$ being the temperature. In contrast, we assume that active fluctuations are an exponentially correlated process
\be
\la \fc_A(x,t) \fc_A(x', t')\ra =  2\Lambda_A \, \frac{e^{-\vert t-t' \vert/\tau_A}}{\tau_A}\delta(x - x') \,. 
\ee
Such a finite correlation time $\tau_A$ in the noise is incompatible with the fluctuation-dissipation relation {(FDR)}~\cite{Basu2008}.  

We choose velocity continuity at the inclusion, and
no flow ($v(-L) =  v(L) = 0$),  at the confining walls.
The local force balance $\ddx \sigma = 0$ implies that the stress is constant $\sigma^{L,R}=\sigma^{L,R}(t)$ within each left and right segments. 
Integrating Eq.\ref{eq:stress1_Maxwell} over the left (L), and the right (R) region leads to
\bea \label{eq:stress}
(\tau_v \p_t + 1) \sigma^{(L)}  &=& \zeta \Delta \mu c_0 + \frac{\eta \dot{X}_t }{L+X_t} + \frac{\int^{X_t}_{-L} (\vartheta_T + \vartheta_A)}{L+X_t} , \, \\
(\tau_v \p_t + 1) \sigma^{(R)}  &=&  \zeta \Delta \mu c_0 - \frac{\eta \dot{X}_t }{L - X_t} + \frac{\int^{L}_{X_t} (\vartheta_T + \vartheta_A)}{L - X_t} , \,
\eea
where $c_0$ is the density of the active stress generators (e.g., actomyosin). Assuming that the turnover of actomyosin is fast, we take its concentration
to be a constant and equal on the left and the right region. 
We now consider the force balance on the point inclusion at $x_t$:
\begin{align} \label{eq:stress}
\sigma^{(L)} - \sigma^{(R)} = f(x_t) - m \dot{v}_t ,
\end{align}
where we have included the inertia of the inclusion and a force  $f(x)$ applied on the inclusion. The velocity of the fluid at the inclusion
 is same as inclusion velocity
$v_t (= \dot x_t)$.
Combining the latter force balance equation with \refn{eq:stress}, we obtain the following dynamics on $\dot x  =v $.
\begin{align} \label{eq:firsteq_Langevin_0}
\tau_v \ddot{v}_t +  \dot{v}_t + \frac{\lambda(x_t)}{m} v_t = 
\frac{f  + \tau_v \dot{f}}{m} + \frac{1}{m}\left[{\theta}_A + {\phi}_T\right], 
\end{align}
where ${\theta}_A = (\int^{X_t}_{-L} \vartheta_A)/(L+X_t) + (\int^{L}_{X_t} \vartheta_A)/(L-X_t)$
and ${\phi}_T$ is defined similarly ($\vartheta_T \leftrightarrow \vartheta_A$), and
\begin{align}
\lambda(x)  = \frac{2 \eta L}{L^2- x^2},
\end{align}
is space-dependent friction term. 
Notice that  $\lambda(x)$ diverges at the boundaries; in the following, wherever necessary we will implicitly assume a regularization $\epsilon$ to temper the divergence,  $\lambda(x) \equiv  2 \eta L/ \l(L+ \epsilon)^2  - x^2 \r$,
 taking the $\epsilon \to 0$ limit at the end. 

The two-point correlations of $\vartheta_A$ and $\vartheta_T$ is
\begin{align} \label{eq:firsteq_Langevin}
\left\langle {\theta}_A(t) {\theta}_A(s) \right\rangle &= 2 \Lambda_A \frac{2 L}{L^2-x^2} \frac{e^{-\lvert t-s \lvert/\tau_A}}{2 \tau_A}, \\
\left\langle {\phi}_T(t) {\phi}_T(s) \right\rangle &= 2 \Lambda_T \frac{2 L}{L^2-x^2} \delta(t-s).
\end{align}
We will make use of the following noise amplitude function 
\begin{align} \label{eq:gA}
\mu^2(x) = \frac{2 L}{L^2-x^2} = \frac{\lambda(x)}{\eta}.
\end{align}
Noticing that $\dot{f} = f^{\prime} v_t$, we define a renormalized time scale $\taumass$ as
\begin{align} 
\frac{1}{\taumass} =  \frac{1}{\tau_m(x_t)} \left(1 - \frac{\tau_v  f^{\prime}}{\lambda(x)}\right),
\end{align}
where $\tau_m(x_t) = m/\lambda(x_t)$. With these redefinitions \refn{eq:firsteq_Langevin_0} reads:
\begin{align} \label{eq:firsteq_Langevin}
\tau_v \ddot{v}_t +  \dot{v}_t + \frac{1}{\taumass} v_t = 
\frac{f +  \mu(x_t) \left[\theta_A + \phi_T\right]}{m} .
\end{align}
In  this work we are interested in the dynamics at time scales larger than the 
the time scale over which inertia of the inclusion gets damped ($\tau_m \sim 10^{-9} \mathrm{s}$  in the cell cortex), the
Maxwell stress relaxation time ($\tau_v \sim 5 \mathrm{s}$, for the cell cortex \cite{Saha2016}), and the active stress correlation time ($\tau_A$ $\sim 60$s), for the cell cortex height fluctuations \cite{Li2014}.
The overdamped Langevin equation obtained by naively setting these time scales to zero leads to, 
\be
\dot x = \f{f}{\lambda} + \f{\mu}{\lambda} \theta_A + \f{\mu}{\lambda} \phi_T, 
\label{eq:naive_overdamped}
\ee
where $\theta_A$, and $\phi_T$ are delta correlated Gaussian noise. We thus see that the confinement gives rise to a multiplicative noise on the dynamics of the inclusion. This arises from an interplay between hydrodynamic interaction, confinement and fluctuations.

As in any situation with multiplicative noise, we must provide an interpretation of the noise term in order to give meaning to \refn{eq:naive_overdamped} \cite{Gardiner, Kampen1981}. The proper choice of convention is dictated by the order of adiabatic elimination of the three time scales. 

In the limit $\tau_m , \tau_A \gg \tau_v$, this reduces to an underdamped Langevin equation, 
\bea
 m \dot{v} + \lambda(x) v =  f(x) +  \mu(x_t) \left[\theta_A + \phi_T\right], 
\eea
where  $\phi_T$ is delta correlated Gaussian noise, and $\theta_A$ is exponentially correlated active noise. This equation has been analyzed in literature \cite{Kupferman2004}, a brief version of the adiabatic elimination of the inertial relaxation and noise correlation times scale is presented in Appendix \ref{ap:limit}. 

As mentioned above, for the cell cortex, the timescale of relaxation of inertia ($\tau_m \sim 10^{-9} \mathrm{s}$) is much smaller compare to the Maxwell time ($\tau_v \sim 5 \mathrm{s}$), and the noise correlation time  ($\tau_n \sim 60 \mathrm{s}$). In the following we derive the overdamped Langevin equation by adiabatically elimination, in the limit $\tau_m \ll \tau_v, \tau_A$.

The long-time solution of \refn{eq:firsteq_Langevin} can be expressed in terms of a Green function $\chi^{(t-s)}_{x_t}$ as 
\begin{align}  \label{eq:newton}
 v_t = \int^{t}_{-\infty} ds \, \chi^{(t-s)}_{x_t}  \f{1}{m}\left\lbrace f   + \mu(x) \left[\theta_A + \phi_T\right] \right\rbrace,
\end{align}
In the following, we justify that in the limit $\tau_m \ll \tau_v$ at first order in $\tau_m $, 
\begin{align} \label{eq:def_chi}
 \chi^{(t-s)}_{x_t} = \sqrt{\frac{\hat \tau_m(x_t)}{\tau_v}} \sin \left[\frac{t-s}{\sqrt{\hat \tau_m(x_t) \tau_v}} \right] \exp\left(-\frac{t-s}{2 \tau_v}\right),
\end{align}
which corresponds to the Green function of \refn{eq:firsteq_Langevin} in the presence of a constant mass time scale. 
The subscript denotes the value of $x$ at which $\hat \tau_m$ is evaluated, and the superscript denotes that the function is of variable $(t-s)$.
We first expand $\tau_m(x_t)$ as:
\begin{align} 
\tau_m(x_t) &= \tau_m(x_s) +  \tau^{\prime}_m (x_t - x_s),
\end{align}
where $x_t - x_s$ can be expressed from \refn{eq:firsteq_Langevin} as
\begin{align} \nn
x_t - x_s &= \frac{1}{m} \int^{t}_{s} dt' \int^{t^{\prime}}_{-\infty} \chi^{(t^{\prime}-s)}_{x_{t^{\prime}}} \left\lbrace f(x_s) + \mu \left[\phi_T + \theta_A \right] \right\rbrace.
\end{align}
where the left-most integral is the variable $t^{\prime}$ and the second on the variable $s$. The latter equation leads to $\tau_m(x_t) = \tau_m(x_s)  + \mathcal{O}(\tau_m^{1/2})$ since $\chi = \mathcal{O}(\tau_m)$ and $\mu = \mathcal{O}(\tau_m^{-1/2})$. At this order, we find that $\taumass(x_t) = \tau_m(x_t)  + \mathcal{O}(\tau_m^{1/2})$. In turn, after expansion of \refn{eq:def_chi} at first order in $\tau_m$, i.e. $\chi_{x_t}^{t-s} = \chi_{x_s}^{t-s}+ d\chi/d\taumass(\taumass(x_t) - \taumass(x_s))$, we find that:
\begin{align}  \label{eq:chi_expansion}
\chi_{x_t}^{t-s} &=\chi_{x_s}^{t-s}  + \mathcal{O}(\tau_m^{3/2}),
\end{align}
which supports our claim that the exact Green function associated to \refn{eq:firsteq_Langevin} converges to \refn{eq:def_chi} at order $\tau_m^{3/2}$.

In the limit $\tau_A \gg \sqrt{\tau_m \tau_v}$, the integral on the continuous functions $f(x)$ and $ \mu(x) \theta_A$ can be simplified as $ \chi^{(t-s)}_{x_t} \sim \tau_m \delta(t-s)$ in the limit $\sqrt{\tau_m\tau_v}  \ll t-s$. However, this relation does not hold over the discontinuous white noise $\mu(x)  \phi_T$. Following \cite{Sancho1982}, we  Taylor expand the noise amplitude $\mu(x)$ to obtain:
\begin{align} 
\mu (x_s) &= \mu(x_t) - \frac{\mu^{\prime} (x_t) \mu(x_t)  }{m} \int^{t}_{s} \!\!\! \int^{t^{\prime}}_{-\infty} \! \! \! 
 \chi^{(t^{\prime}-t^{\prime \prime})}_{x_t} \phi_T(t^{\prime \prime}), \label{eq:mu_expansion}
\end{align} 
which holds up to $ \tau^{1/2}_m $ terms. From Eqs. (\ref{eq:newton}) and (\ref{eq:mu_expansion}), we obtain the following Langevin equation that is valid at $\tau^{3/2}_m$ order:
\begin{align}
v_t = \frac{\tau_m}{m}  \left\lbrace f +  \mu \left[ \theta_A  + \theta_T \right] \right\rbrace  + \Omega_T,   \label{eq:velocitysdeorder3} 
\end{align}
where  
\begin{align}
\theta_T= \frac{1}{\tau_m} \int^{t}_{-\infty} ds \,  \chi^{(t-s)}_{x_s}  \phi_T(s)
\end{align}
 is colored noise  and
\begin{align}
\Omega_T &= \frac{\mu \mu^{\prime}}{m^{2}} 
\int^{t}_{-\infty} \chi^{(t-t^{\prime})}_{x_t} \phi_T(t^{\prime})
 \int^{t^{\prime}}_{t} \!
\int^{t^{\prime \prime}}_{-\infty} \, \chi^{(t^{\prime \prime} -t^{\prime \prime \prime})}_{x_{t^{\prime \prime}}}  \phi_T(t^{\prime \prime \prime}). \nonumber  
\end{align}

We first present a series of identities which will be useful in the forward calculation. Based on \refn{eq:def_chi}, we find that
\begin{align} 
\int^{t_1}_{-\infty} \! dt^{\prime} \chi_x^{t_1-t^{\prime}} \chi_x^{t_2-t^{\prime}} 
&\underset{\tau_m \ll \tau_v}{\sim} \frac{\tau_m}{2} \cos\left(\frac{t_1-t_2}{\sqrt{\tau_m \tau_v}} \right) e^{-\lvert t_1-t_2 \rvert/(2 \tau_v)},
\label{eq:identity2}
\end{align}
which converges to $\tau^{2}_m \delta(t_1-t_2)$ in the limit $\sqrt{\tau_m \tau_v} \ll (t_1-t_2)$ and 
\begin{align}
\int^{t}_{-\infty} \! dt^{\prime} \chi_x^{t-t^{\prime}} \int^{t}_{t^{\prime}} \! dt^{\prime \prime} \chi_x^{t^{\prime \prime}-t^{\prime}} &= \frac{8 \tau_m^2 \tau_v^2}{(\tau_m +4 \tau_v)^2},  \label{eq:identity1}
\end{align}
which converges to $\tau_m^2/2$ in the limit $\tau_m \ll \tau_v$.
Secondly, we notice that:
\begin{align} \label{eq:exp_exp}
 \int^{t_1}_{-\infty} dt^{\prime} \int^{t_2}_{-\infty} dt^{\prime \prime}  \, 
\frac{e^{-\frac{t_1-t^{\prime} + t_2 - t^{\prime \prime}}{\tau_A}} }{\tau^{2}_A} 
\delta_{t^{\prime},t^{\prime \prime}} &= \frac{e^{-\frac{\lvert t_2-t_1 \lvert}{\tau_A}}}{2 \tau_A},
\end{align}
which converges to $\delta(t_2-t_1)/2$ in the limit $\tau_A \ll (t_2-t_1)$.
Finally, we notice that
\begin{align}
&\int^{t_2}_{-\infty} dt^{\prime} \int^{t_1}_{-\infty} dt^{\prime \prime}  \, 
\frac{e^{-(t_2-t^{\prime})/\tau_A} }{\tau_A} 
\chi_{x}^{t_1-t^{\prime}} \nonumber
\delta_{t^{\prime},t^{\prime \prime}}  \\
&= \int^{\min(t_2,t_1)}_{-\infty} dt^{\prime} \, \frac{e^{-(t_2-t^{\prime})/\tau_A} }{\tau_A}
\chi_{x}^{t_1-t^{\prime}},
\end{align}
can be further simplified using the following results:
\begin{align}
\int^{t_1}_{-\infty} dt^{\prime} \, \frac{e^{-(t_2-t^{\prime})/\tau_A} }{\tau_A}
\chi_{x}^{t_1-t^{\prime}}
&\underset{\tau_m \ll \tau_v}{\sim}
\tau_m \frac{e^{-\frac{t_2-t_1}{\tau_A}}}{\tau_A}, \label{eq:exp_chi_1}
\end{align}
which converges to $\tau_m \delta(t_1-t_2)$ in the limit $\tau_A \ll (t_2-t_1)$ while:
\begin{align}
\int^{t_2}_{-\infty} dt^{\prime} \, \frac{e^{-\frac{t_2-t^{\prime}}{\tau_A}}}{\tau_A}
\chi_{x}^{t_1-t^{\prime}}
&\underset{\tau_m \ll \tau_v}{\sim} \frac{\tau_m}{\tau_A} \cos\left( \frac{t_1 - t_2}{{\sqrt{\tau_m \tau_v}}} \right)e^{- \frac{t_1-t_2}{2 \tau_v}},\label{eq:exp_chi_2}
\end{align}
which converges to $(2 \tau^2_m)/\tau_A \delta(t_1-t_2)$ in the limit $\sqrt{\tau_m \tau_v} \ll (t_2-t_1)$.

We now compute the average of \refn{eq:velocitysdeorder3} over the realizations of the processes $\theta_T$ and $\theta_A$ (at a constant $x_t$).  We notice that $\left\langle \phi_T(t^{\prime}) \theta_A(t^{\prime \prime}) \right\rangle = 0$ -- due to the independence of the thermal and active noise sources -- and we use the identity from \refn{eq:identity1} to show that 
\begin{align} \nonumber
\left\langle \Omega_T \right\rangle &= \frac{\mu \mu^{\prime}}{m^{2}} 
\int^{t}_{-\infty} dt^{\prime} \, \chi_{x}^{t-t^{\prime}} 
\int^{t^{\prime}}_{t} \! dt^{\prime \prime} 
\int^{t^{\prime \prime}}_{-\infty} dt^{\prime \prime \prime} \, \chi_{x}^{t^{\prime \prime} -t^{\prime \prime \prime}}
\delta_{t^{\prime},t^{\prime \prime \prime}}, \\
&= 
- \frac{\tau^2_m}{2 m^{2}} \mu^{\prime}_T  \mu_T.
\label{eq:multiplicative_noise_expanded}
\end{align}
Similar steps of calculations leads to $\left\langle \Omega_T(t) \Omega_T(s) \right\rangle = 0$ at order $\tau^{3/2}_m$. The two-time correlation of $\phi_T$ can calculated using \refn{eq:identity1}:
\begin{align}
\left\langle \theta_T(t_1)\theta_T(t_2) \right\rangle &= \frac{1}{\tau_m^{2}} 
\int^{t_1}_{-\infty} \! \chi_{x_{t_1}}^{(t_1 - t^{\prime}_1)}
\int^{t_2}_{-\infty} \! \chi_{x_{t_2}}^{(t_2 - t^{\prime}_2)} 
\, \delta_{t^{\prime}_1,t^{\prime}_2}, \nn \\
&=\frac{1}{2 \tau_m} \cos\left(\frac{t_1-t_2}{\sqrt{\tau_m \tau_v}} \right) e^{-\lvert t_1-t_2 \rvert/(2 \tau_v)}, \nn 
\end{align}
which converges to  $ \delta(t_1-t_2)$ in the limit of a long observation time. This leads to the following dynamics at order $\tau^{3/2}_m$
\begin{align} \label{eq:FKf}
\dot{x}_t = \frac{f(x)}{\lambda(x)}
- \frac{\Lambda_T \mu(x) \mu^{\prime}(x) }{2 \lambda^{2}(x)} 
+ \frac{\mu(x) \left[\theta_A 
+ \theta_T\right]}{\lambda(x)},
\end{align} 
where $\theta_{A/T}$ are noise sources that are correlated in time and of equal time variance $\Lambda_{A,T}$.  In the limit $\tau_v$, and $\tau_A$ going to zero, these colored noise sources go to a white noise source, and hence in the white noise limit \refn{eq:FKf} should be interpreted in Stratonovich conventions \cite{Gardiner}.
 
The Fokker-Planck equation associated to \refn{eq:FKf} reads: 
\begin{align} \label{eq:fokker_planck_1}
\pa_t  P &= \f{\p}{\p x}  \l -\frac{f(x) }{\lambda(x) }  + \frac{\Lambda_T \mu(x) \mu^{\prime}(x)}{2 \lambda^2(x) } + \frac{\Lambda \mu(x)}{2 \lambda(x) } \partial_x  \frac{\mu(x)}{\lambda(x) }  \r P,
\end{align}
where $\Lambda = (\Lambda^2_T + \Lambda^2_A)^{\sfrac{1}{2}}$. 

In the absence of an active noise ($\Lambda_A=0$), we find that the steady state solution of \refn{eq:fokker_planck_1} is the Boltzmann distribution $P  \propto \exp(-U(x)/k_b T)$, where $f = - \pa_x U$ and  $k_b T = \Lambda_T/\eta$.

The dynamics  
\be
\dot{x} = \f{f(x)}{\lambda(x)} + \f{\mu(x)}{\lambda(x)} \theta_T, 
\ee
 is consistent with thermodynamics only when the Gaussian white noise is interpreted with the Hanggi-Klimontovich convention \cite{Lau2007}

Whenever $\Lambda_A>0$, the steady state of  \refn{eq:fokker_planck_1} is not Boltzmann distributed and there is no effective temperature such that $P  \propto \exp(-U(x)/k_b T_{\mathrm{eff}})$. Remarkably, \refn{eq:fokker_planck_1} shows that, in general, the coexistence of thermal and active noise sources cannot be described by any $\alpha$-convention \cite{Lau2007}, e.g. neither by the Stratonovich ($\alpha = 1/2$) nor by the Hanggi-Klimontovich ($\alpha = 1$) conventions.

When the thermal noise is negligible ($\Lambda_T = 0$), \refn{eq:fokker_planck_1} amounts to
\begin{align} \label{eq:Strato}
\partial_t P &=  \partial_x \left\lbrace - \frac{f}{\lambda} P  + \Lambda_A \frac{\mu}{2 \lambda} \partial_x  \frac{\mu}{\lambda}  P  \right\rbrace,
\end{align}
which corresponds to the Stratonovich convention of \refn{eq:FKf}.
We will come back to the analysis of this equation in Section \ref{sec:rigid}.

To summarize, in the overdamped equation we naively obtained by setting the timescales to zero, 
\be
\nn \dx =   f(x)\f{(L^2 - x^2)}{2\eta_c L}  + \sqrt{\frac{\Lambda_c}{\eta_c^2}\f{(L^2 - x^2)}{\tilde L}} \l \theta_A + \theta_T\r,
\ee
The thermal noise should be interpreted in Hanggi-Klimontovich convention \cite{Hanggi1982, Klimontovich1990}, and the active noise should be treated in Stratonovich convention \cite{Stratonovich, Gardiner}. 
In the rest of the paper this is the convention that will be used. 

\section{Extended inclusion in active fluid} \label{sec:basicequation}
\begin{figure}
   \centering
   \includegraphics[width=80mm]{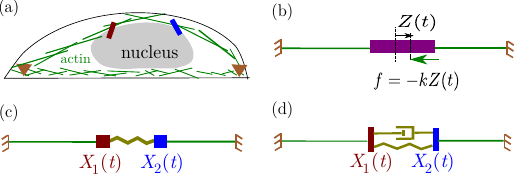}
   \caption{
   (a) Schematic of the cell nucleus embedded in an active cytoplasm and confined by actin stress fibers. Following \cite{Rupprecht2017}, we model the nucleus as an    inclusion embedded in an active fluid confined by the cell boundary. To address various possible contexts, we treat the inclusion in turn as (b) a passive rigid element (Sec.\,\ref{sec:rigid}),
   (c) an elastic element (Sec.\,\ref{sec:elastic}) and (d) a viscoelastic (Kelvin-Voigt) element (Sec.\,\ref{sec:viscoelastic}).
   } 
	\label{fig:schematic}
   \end{figure}

Having established our convention, we 
will now present the general set of equations valid for extended inclusions of any type in 1d.
The context that we will focus on is the position and shape of the nucleus \cite{Rupprecht2017} confined within the cell (Fig.\,\ref{fig:schematic}(a)). We now treat the active gel surrounding the inclusion and confined by the cell boundary as an active Stokes fluid (Fig.\,\ref{fig:schematic}(a)); we are thus interested
in the dynamics at time scales larger than the Maxwell stress relaxation time ($\tau_v \sim 5 \mathrm{s}$, for the cell cortex)~\cite{Saha2016}
and the time scale over which inertia of the inclusion gets damped ($\tau_m \sim 10^{-9} \mathrm{s}$). The active Stokesian
fluid comprises a collection of stochastic force dipoles -  characterising the statistics of remodelling  of actomyosin
- represented by an active noise that is correlated over a  finite time.
In our treatment, we study the stochastic dynamics over time scales larger than the active stress correlation time $\tau_A$ ($\sim 60$s) \cite{Li2014}.
 
Consider an inclusion $(I)$ embedded in a 1d active fluid confined between hard walls at $x=\pm L$ (Fig.\,\ref{fig:schematic}(b)-(d)). We denote the edges of the inclusion as
 $x_1(t) \leq x_2(t)$ and the viscous fluid regions to the left and right of the
inclusion as $(L)$ and $(R)$, respectively.

The hydrodynamic variables describing the bulk are the actomyosin concentration $c$ and the hydrodynamic velocity $v$. 
The velocity field of this active Stokesian fluid is determined by local force balance, $\ddx \sigma(x,t) = 0$. We express the local stress given by \refn{eq:stress1_Maxwell}, which for time scales larger than $\tau_v$ reduces to, 
\be
\sigma^{L,R} = \eta_c \ddx v - \zeta \m c + \fc .
\label{eq:stress1}
\ee
For convenience, we study the effects of thermal and active noise separately. When the noise is active, in which case the thermal noise can be ignored, and when active noise is zero, in this case only thermal noise is present.  Henceforth, we will denote the noise as $\fc$, with variance $\Lambda_c$,  irrespective of whether the noise is thermal or active, with the
understanding that $\Lambda_c = k_b T \eta_c$ for the thermal case.

The inclusion will be  treated in turn as a passive rigid element, a passive elastic element and a viscoelastic (Kelvin-Voigt) element.
We denote the bulk stress in the inclusion as $\sigma^{I}(t)$, whose form we discuss in subsequent sections. In general, there could also 
be a stress 
at the boundary between the inclusion and the embedding active fluid, 
arising for instance from a 
confining force, $f(x)$, 
which favors a centering of the nucleus. This could arise from a variety of sources, such as confinement due to microtubules and motors \cite{Dupin2011, Gundersen2013, Morris2002},  or the
 geometry of stress fibers constraining the nucleus \cite{Versaevel2012, Rupprecht2017}. 

We now specify the boundary conditions both at the edges of the inclusion and the rigid confining walls. 
We choose velocity continuity at the boundaries of the inclusion $v^{I}(x_1) = v^{L}(x_1)$ and $v^{I}(x_2) = v^{R}(x_2)$, and either
{\it no flow} $v(-L) = v(L) = 0$ or {\it finite flow}  $v(-L) =  v_L, v(L) = -v_R$,  at the confining walls.

The local force balance $\ddx \sigma = 0$ implies that the stress is constant $\sigma=\sigma(t)$ within each left and right segments. Integrating Eq.\,\ref{eq:stress1} over $x$ in the regions $(L)$ and $(R)$, we obtain that
\bea
\label{eq:inclusiondynam_1}
\eta_c \dx_1 &=& \eta_c v_L + (x_1 +  L) \sigma^L(t) + \zeta \Delta \mu c_{L} -  \int^{x_1}_{-L} \fc \,dx, \\
\nn \eta_c \dx_2 &=& -\eta_c v_R + (x_2 - L) \sigma^R(t) - \zeta \Delta \mu c_{R} +  \int^{L}_{x_2} \fc \,dx,\\
\label{eq:inclusiondynam_2}
\eea
where $c_{L} = \int_{-L}^{x_1} c(x) dx$ and $c_{R} = \int^{L}_{x_2} c(x) dx$ are the  total bulk concentrations of actomyosin.
We 
 analyse the limit when the turnover of actomyosin is fast, this gives rise to a  constant {\it local actomyosin density} $c_0$, thus, $c_{L} = (x_1 + L) c_0$ and $c_{R} = (L - x_2) c_0$.
In Appendix \ref{ap:constant_c}, we discuss the limit of slow actomyosin  turnover, when  the  {\it total actomyosin number} $c_{L}, c_{R}$ can be taken to be constant.
In the fast turnover limit, Eqs.\,\ref{eq:inclusiondynam_1} and \ref{eq:inclusiondynam_2} become to
\bea
\label{eq:x1}
\nn \eta_c \dx_1 &=& (x_1 +  L) (\sigma^{L}(t) + \zeta \Delta \mu c_0) + \eta_c v_{L} -  \sqrt{x_1 + L}  \, \theta_1, \\
\\
\nn \eta_c \dx_2 &=& (x_2 - L)( \sigma^{R}(t)  + \zeta \Delta \mu c_0) - \eta_c v_{R} + \sqrt{L - x_2}  \, \theta_2,
\\
\label{eq:x2}
\eea
where $\theta_1$ and $\theta_2$ are Gaussian white noise of variance $2\Lambda_c$ which originates from the spatial integration of the fluctuating stress: $\int_{-L}^{x_1}\fc\, dx = \sqrt{L + x_1} \, \theta_1(t)$
and $\int_{x_2}^{L}\fc\, dx = \sqrt{L - x_2} \, \theta_2(t)$.

We point out that Eqs.\,\ref{eq:x1}-\ref{eq:x2} relate the coordinates of the inclusion edges $x_1$ and $x_2$ to the bulk stresses $\sigma^L$ and $\sigma^R$. These bulk stresses are determined by the bulk flows generated by the active stress and the stress in the inclusion $(I)$. These are related by stress continuity,
$\ddx \sigma(x,t) = 0$ :
 $\left[\sigma^L(t) - \sigma^{I}(t)\right] = f(x_1)$ and  $\left[\sigma^R(t) - \sigma^{I}(t)\right] = -f(x_2)$, where $f$ is the confining force acting on the boundaries of the inclusion.
 

Here we analyze the equations in presence of one noise source, however the formalism can be easily extended to include multiple noise sources (see,  Appendix \ref{ap:multiple_noise}).

%

\section{Rigid inclusion in an active fluid} \label{sec:rigid}

We first model the inclusion as a rigid object of fixed size $x_2 - x_1 =2l$; being rigid, the confining force applied at the boundary of the inclusion can be transferred to the 
centre-of-mass (COM) coordinate $x_{cm}\equiv 2z \equiv (x_1+x_2)/2$ (see also Fig.\,\ref{fig:schematic}(b)). Now the stresses are related by $\sigma^L(t) =  \sigma^R(t) + f(x_{cm})$. 

Combining Eqs.\,\ref{eq:x1} and \ref{eq:x2}, we find that the dynamics of the COM coordinate reads
\bea
\nn \dx_{cm} = - \f{v_0}{ 2\tilde L} x_{cm}  &+&   f(x_{cm})\f{(\tilde L^2 - x_{cm}^2)}{2\eta_c \tilde L}  \\
&+& \sqrt{\frac{\Lambda_c}{\eta_c^2}\f{(\tilde L^2 - x_{cm}^2)}{\tilde L}}\theta,
\label{eq:rigid1}
\eea
where we define the length $\tilde L= L - l$, the velocity $v_L = - v_R = v_0$ and the Gaussian white noise of unit variance $\theta$.

This equation not surprisingly, is same as that of point inclusion obtained in Section \ref{sec:noise_limit}, when $\tilde L$ is identified with $L$.
As noted above, this Langevin dynamics with multiplicative noise arises as an interplay between hydrodynamic interaction, confinement and fluctuations. 

The $\tilde L^2 - x_{cm}^2$ term above is a consequence of the long range hydrodynamic interaction between the confining wall and the inclusion, which allows the effect of the
boundary to be felt far into the bulk.

The Langevin equation, for both thermal and active cases, is of the general form, $\dot x = F(x)  + G(x)\theta(t)$, for which the corresponding Fokker-Planck equation is \cite{Lau2007},
\be \label{eq:fokkerplanck}
\nn \ddt P = \ddx \l -F(x)  - \alpha G(x) G'(x) + \frac{1}{2} \ddx G^2(x) \r P,
\ee
where $\alpha = 1$ (Hanggi-Klimontovich) corresponds to the thermal case and $\alpha = 1/2$ (Stratonovich)  corresponds to the active noise (see Appendix A).

\begin{figure}
   \centering
   \includegraphics[width=80mm]{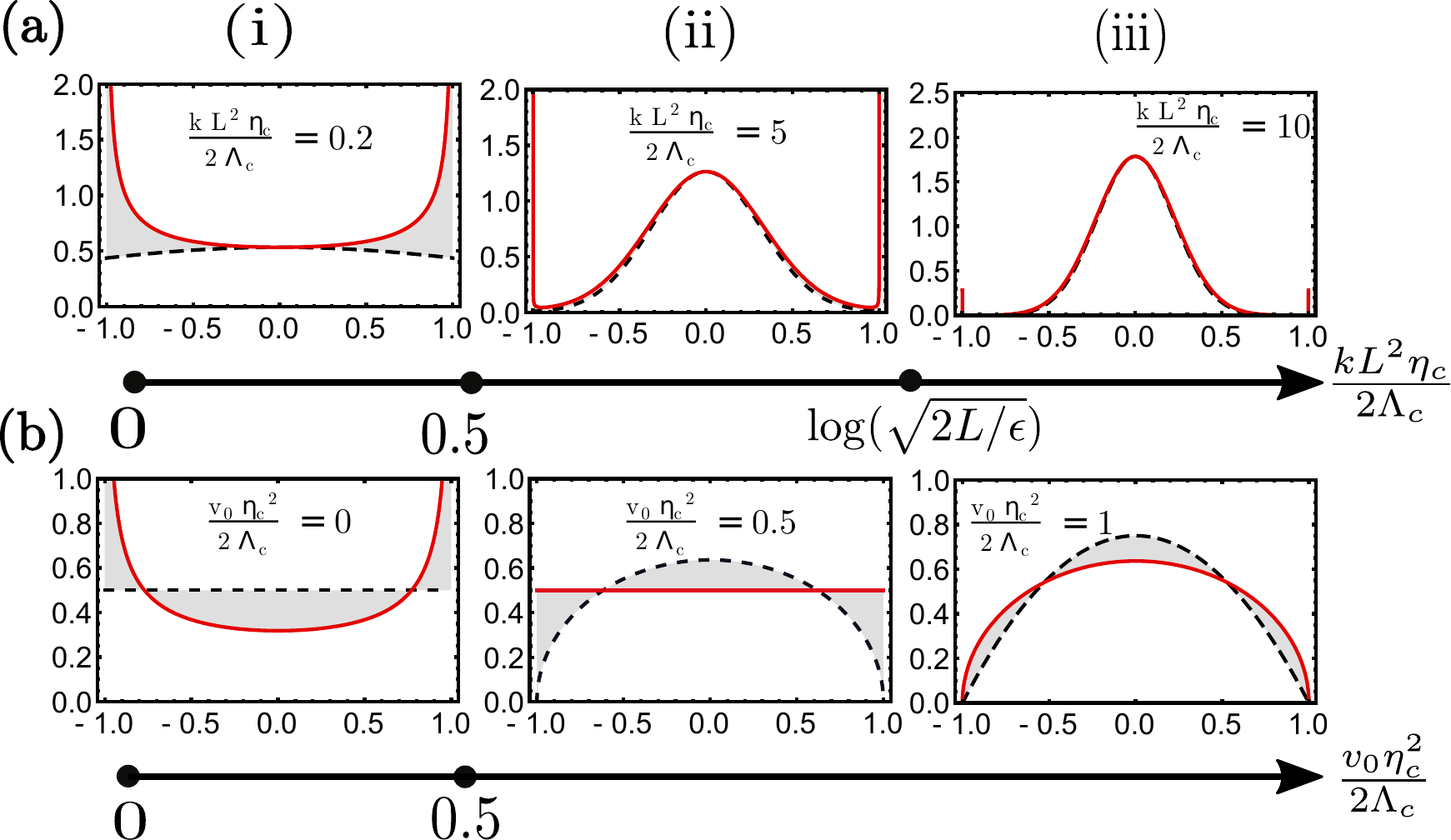}
   \caption{(color online) Steady-state probability distribution $P(x_{cm})$ for the inclusion center of mass (COM) $x_{cm}$ as a function of (a) the dimensionless stiffness of the confining potential $(k L^2 \eta_c)/(2\Lambda_c)$ and (b) of the dimensionless boundary flow $(v_0 \eta_c^2 )/(2\Lambda_c)$. (a) For an active noise ($ \alpha = 1/2$; indicated by a solid red line), we distinguish between three regimes of
   (i) boundary adhesion, (ii) boundary adhesion with metastable centering and (iii) stable centering. The existence of the boundary adhesion phase in the active situation contrasts with the two phases found in the thermal situation ($\alpha = 1$, dashed black line): a uniform distribution and a stable centering. (b) The active case, $ \alpha = 1/2$ (solid red line), shows a transition from  (i) boundary adhesion  to (iii) stable centering via  (ii) a uniform distribution at $0.5$. }
	\label{fig:rigidpx}
   \end{figure}

Following \cite{Gardiner}, we find that the steady state solution corresponding to Eq.\,\ref{eq:fokkerplanck} reads
\be \label{eq:general_steady}
P(x) = N (G(x))^{2(\alpha - 1)}\exp \l \int \frac{2F(y)}{G^2(y)}dy \r, 
\ee
where $N$ is a normalization constant. To derive \ref{eq:general_steady}, we assume that the inclusion cannot exit the confining domain and we consider that there is no flux for the inclusion at $x=\pm L$. After identification of the functions $F$ and $G$ from Eq.\,\ref{eq:rigid1}, Eq.\,\ref{eq:general_steady} becomes
\bea \label{eq:form2}
P(x_{cm}) &=& N e^{ - \f{\eta_c}{\Lambda_c}(V(x_{cm}) + V_{\alpha}(x_{cm}) + U(x_{cm}))},
\eea
where $N$ is a normalization constant and
\begin{enumerate}[(i)]
\item $V(x_{cm}) = \int^{z}_{0} f(u)\, du$ is the confining 
 mechanical potential,
\item $V_{\alpha}(x_{cm}) = (1 - \alpha ) (\Lambda_c/\eta_c) \, \log (\tilde L^2 - x_{cm}^2)$ corresponds to an effective potential which represents the contribution from the multiplicative noise, 
\item $U(x_{cm}) = (v_0 \eta_c/2)  \log (\tilde L^2 - x_{cm}^2)$ corresponds to an effective potential which represents the contribution from the boundary flow.
\end{enumerate}
In the following, we show that the interplay of these contributions gives rise to sharp transitions in the shape of the steady state distribution, as displayed in Fig.\,\ref{fig:rigidpx} for $V(x_{cm}) = k x_{cm}^2/2$.

To identify the origin of these transitions, it is useful to look at the explicit forms of the distribution in the absence of boundary flow $v_0 = 0$, and confining potential $V(x_{cm}) = 0$.
In the case of thermal fluctuations ($\alpha = 1$) the noise-induced potential vanishes, thus the steady state probability distribution is 
 flat (ie. $P(x_{cm}) =1/(2\tilde L)$), in agreement with the Boltzmann distribution in a flat energy landscape.  

   \begin{figure}
   \centering
   \includegraphics[width=80mm]{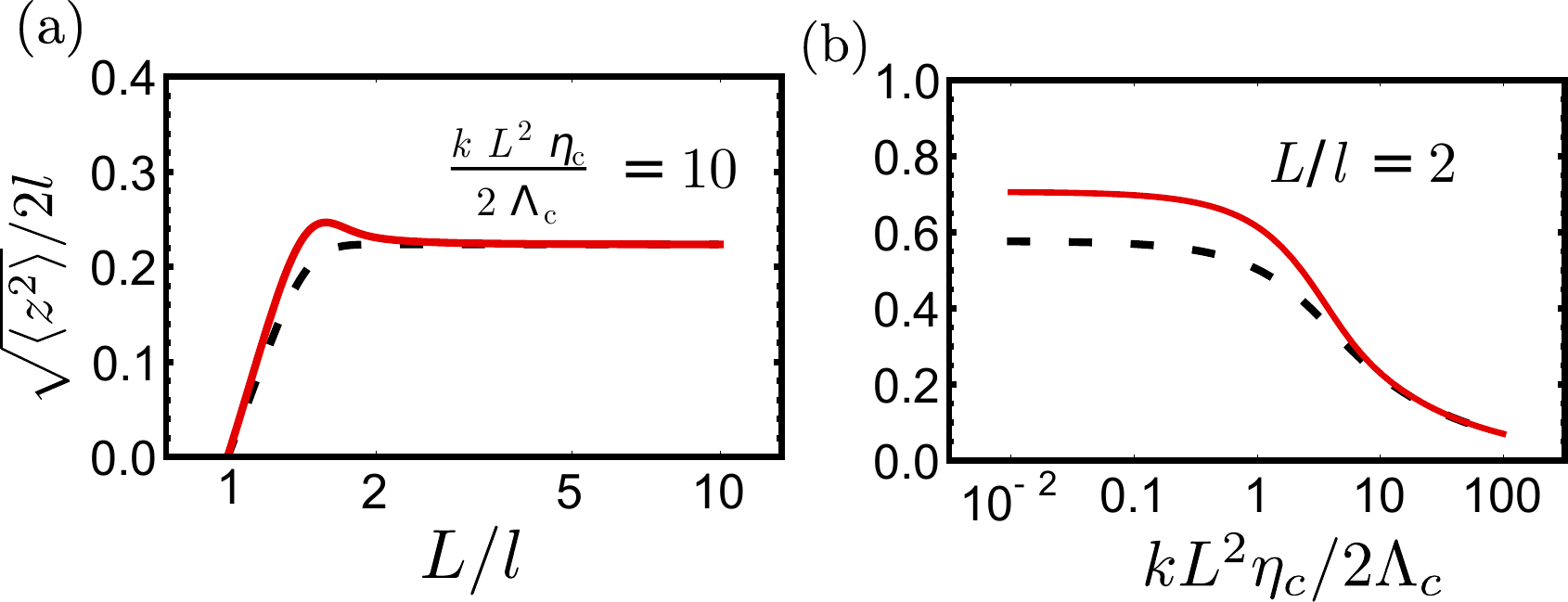}
   \caption{Variance of  the dimensionless COM position $\sqrt{\langle x_{cm}^2\rangle}$  for thermal (dashed black line) and active (solid red line) noises, with $v_0 = 0$ as a function of  (a) system  size  $L$, where it goes from being
  strongly affected by the boundary, when $L$ is small, to reaching a limiting value determined by the confining strength for large $L$; and 
   (b) confining strength $(k L^2 \eta_c)/(2\Lambda_c)$, where it asymptotes to a fixed value for small stiffness and goes to zero for large stiffness.
  	    } 
	\label{fig:rigidx2}
   \end{figure}

On the other hand, when the noise is active, $\alpha = 1/2$, 
 \be
 P(x_{cm}) = \f{\pi}{\sqrt{2(\tilde L^2 -  x_{cm}^2)}},
 \label{eq:formdiv}
\ee
Notice that the distribution Eq.\,\ref{eq:formdiv} diverges at the edges $x = \pm L$; this corresponds to a higher preference of the  inclusion to be at the boundaries (see Fig.\,\eqref{fig:rigidpx}). The inclusion de-centering occurs although there is no net mechanical
 force applied to inclusion; this is solely caused due to the noise-induced effective potential $V_{\alpha}$. 

We point out that, in a more general situation, the divergence can be even stronger than in Eq.\,\ref{eq:formdiv} and can prevent the distribution from being normalizable. This situation occurs under the condition that $(v_0 \eta_c^2)/(2 \Lambda_c)  + ( \alpha - 1)  \leq -1$.
Under this condition, depending on the initial condition the inclusion is then fixed at one of the two boundaries at steady state.

We now include the contribution from the confining potential which we assume to be harmonic: $V(x_{cm}) = k x_{cm}^2/2$, keeping $v_0 = 0$. We show that the competition between the confining potential -- which favors centering the nucleus -- and the
 noise-induced effective potential -- which induces adhesion to the cell boundary -- leads to a sharp transition as
the strength of harmonic potential $k$ or the magnitude of the boundary flow $v_0$ are varied (Fig.\,\eqref{fig:rigidpx}).
This can be thought as an active fluctuation-induced wetting-dewetting transition \cite{DeGennes1985}.

Since the distribution from Eq.\,\ref{eq:formdiv} diverges on the two edges, the probability is maximum at the boundary for all values of $k$. So to compute the phase diagram for harmonic confinement, we regularize it using a cutoff distance $\epsilon$ from the boundary, setting its value in this boundary region to be $\int_{L - \epsilon}^{L}P(x_{cm})\,dx_{cm}/\epsilon$. The transition point $\log(\sqrt{2L/\epsilon})$ in  Fig.\,\eqref{fig:rigidpx} is the value of $(k L^2 \eta)/(2\Lambda)$ for which this boundary value equals $P(0)$. 

Based on Eq.\,\ref{eq:form2}, and setting boundary flow $v_0 = 0$, we find that the variance of COM can be expressed in terms of the hypergeometric $_1{F}_1$ and Gamma functions \cite{Arfken},
\be \label{eq:variance_of_COM}
\la x_{cm}^2\ra = \frac{\tilde L^2}{2} \left[\frac{\, _1{F}_1\left(\frac{3}{2};(1 - \alpha) +\frac{5}{2};-k \tilde L^2\right)\Gamma \l (1 - \alpha) + \f{3}{2} \r}{\, _1{F}_1\left(\frac{1}{2};(1 - \alpha) +\frac{3}{2};-k \tilde L^2\right)\Gamma\l (1 - \alpha) + \f{5}{2} \r} \right].
\ee
We represent the behavior of Eq.\,\ref{eq:variance_of_COM} as a function both the system size $L$ of the stiffness of the confining potential $k$ (Fig.\,\eqref{fig:rigidx2}). In the limit $k = 0$, the latter expression simplifies into $\la x_{cm}^2\ra = \tilde{L}^2/2$, when fluctuations are active, and to $\la x_{cm}^2\ra = \tilde L^2/3$ when fluctuations are thermal.

Of course, in a realistic context, one needs to include an {\it inward pressure}
from the compressed active components in the cytoplasm arising from steric hindrance etc., however the symmetry broken de-centering of the steady state 
distribution due to activity, would still persist.

The fluctuation induced interaction discussed above is robust, and should arise whenever the noise has a multiplicative nature.  The multiplicative nature of the noise resulting from field fluctuations integrated over space is not specific to one dimension and should 
 hold in higher dimensions. For instance, the dynamics of a spherical colloid in a 3-dimensional fluctuating incompressible fluid confined between two parallel walls is 
 described by a Langevin equation \cite{Lau2007} that is identical to Eq.\,\ref{eq:rigid1} with $v_0 = 0$. However \cite{Lau2007} deals with thermal fluctuations, while in addition we study the effects of non-thermal fluctuations.

\section{Elastic inclusion in an active fluid} \label{sec:elastic}

We next model the inclusion as a passive linear elastic element of unloaded length $2l$, with a stress given by
$\sigma^I = B\partial_x u$, where $u$ is the strain from the unloaded configuration and $B$ is the elastic modulus of the inclusion. 
Similar to Sec.\,\ref{sec:rigid}, the local force balance implies that the bulk stresses are constant, eg. $\sigma^I=\sigma^I(t)$. Hence, integrating $\sigma^{I}$ from $z/2 -l$ to $z/2 + l$, we find that the stress along the elastic inclusion reads
\be
 \sigma^I(t) =  \f{B}{2l} \l y - 2l \r,
\ee
where $y - 2l = \int^{l}_{-l} dx \partial_x u(x)$ is the extension of the inclusion. We first derive the general Langevin equation for the inclusion dynamics, before discussing the results for thermal and active fluctuations.

Continuity of stress across the inclusion boundary in the presence of a confining force $f = - \p_x V$ implies $\sigma^{L}(t) = \sigma^{I}(t) + f(x_1)$ and  $\sigma^{R}(t) = \sigma^{I}(t) - f(x_2)$. After substitution in Eqs.\,\ref{eq:x1},\ref{eq:x2}, we obtain
\bea
\label{eq:passiveinclusion}
\nn \eta_c \dx_1 &=& \eta_c v_L + \frac{B}{2l} (x_1 +  L)(y - y_0) - f(x_1)(x_1 + L) \\
&-&  \sqrt{L+x_1}\, \theta_1,  \\
\nn \eta_c \dx_2 &=& -\eta_c v_R + \frac{B}{2l} (x_2 - L) (y - y_0) + f(x_2)(x_2 - L)  \\
&+&   \sqrt{L - x_2}\, \theta_2,
\eea
where we define the length $y_0 = 2l(1 - \zeta \Delta \mu c_0/B)$. We perform a change of variable to express Eq.\,\ref{eq:passiveinclusion} in terms of the vector ${\bf X } =  \begin{bmatrix} y & z \end{bmatrix}^T$, where $y \equiv x_2 - x_1$ is the inclusion length and $z \equiv x_1 + x_2$ is now twice the COM coordinate, to obtain the following multivariate Langevin equation,
\be
\dot{{\bf X}} =  {\bf F} + {\bf G} \cdot {\bf \theta},
\label{eq:multivar}
\ee
where we define the drift vector {\bf F} as
\begin{widetext}
\be \label{eq:force_vector}
{\bf F} = \frac{1}{\eta_c}\begin{bmatrix} \frac{B}{2l}(y - y_0) (y - 2L) - \eta_c v_0 + f(x_1)(x_1 + L) + f(x_2)(x_2 - L) \\   \frac{B}{2l}(y - y_0) z - f(x_1)(x_1 + L) +f(x_2)(x_2 - L) \end{bmatrix},
\ee
\end{widetext}
and the matrix 
\be \label{eq:G_matrix}
{\bf G} = \sqrt{\f{2\Lambda_c}{\eta_c^2} }\begin{bmatrix} \sqrt{L + x_1} &    \sqrt{L - x_2} \\  - \sqrt{L + x_1} & \sqrt{L- x_2} \end{bmatrix} ,
\ee
which operates on the Gaussian white noise vector ${\bf \theta}  = \begin{bmatrix} \theta_1 & \theta_2\end{bmatrix}^T$. For convenience, we also introduce a diffusion matrix
\bea \label{eq:diffusion_matrix}
{\bf D} &=& {\bf G} \cdot {\bf G}^T = \frac{2 \Lambda_c}{\eta_c^2}\begin{bmatrix} 2L - y &   - z \\  - z & 2L-y \end{bmatrix}  .
\eea

The Fokker-Planck equation corresponding to the Langevin Eq.\,\ref{eq:multivar} reads
\bea \label{eq:fokkerplanck_general}
\nn \partial_t P &= & \frac{\partial}{\partial x_i} \l - F_i  - \alpha \f{\partial G_{ik}}{\partial x_j} G_{jk}+ \frac{1}{2} \frac{\partial}{\partial x_j} G_{ik}G_{jk} \r P,
\eea
where we recall that $\alpha = 1$ corresponds to the thermal case and $\alpha = 1/2$ to the active case \cite{Lau2007}.
Following \cite{Gardiner}, we introduce the potential vector
\be \label{eq:potential_condition}
\nn H_i \equiv \partial_{i} \log P =   D_{ik}^{-1} \l 2 F_k  + 2\alpha \f{\partial G_{kl}}{\partial x_j} G_{jl} -  \f{\p}{\p x_j} D_{kj} \r,
\ee
where we consider that repeated latin indices are summed over. Based on Eqs.\,\ref{eq:force_vector}-\ref{eq:diffusion_matrix},
we check that the potential condition 
\bea
\f{\p H_i}{\p x_j} = \f{\p H_j}{\p x_i},
\label{eq:potentialcondition}
\eea
is satisfied. Therefore, the steady state distribution can be expressed as $P(y,z) = N e^{- \phi}$, where the effective potential $\phi$ reads:
\bea \label{eq:steady_potential}
\nn \phi &= & \f{\eta_c}{ \Lambda_c} \l \f{B}{4l}(y - y_0)^2 + V(z +y) + V(z -y) \r\\
&+& \l -\f{ v_0 \eta_c^2}{2\Lambda_c} + (1 - \alpha)\r \log\l(2L - y)^2 - z^2 \r.
\eea
As in the rigid inclusion considered in Sec.\,\ref{sec:rigid}, the effective potential $\phi$ is built from the confining potentials
$V(z\pm y)$ (at $x_2$, $x_1$, respectively) and the  fluctuation-induced potential $V_{\alpha} =  (1 - \alpha) \log\l(2L - y)^2 - z^2\r$.
In addition, the effective potential has a contribution from the elastic energy of the inclusion, $B (y - y_0)^2/(4l)$.

In the case of a harmonic confining potential $V(x_1,x_2) =-k (x^2_1 - x^2_2)$, the potential reads:
\bea
\nn \phi &= & \f{\eta_c}{ \Lambda_c}\l \l \f{B}{4l} + \f{k}{4} \r(y - \tilde y_0)^2 + \f{k}{4}z^2  \r\\
&+& \l -\f{ v_0 \eta_c^2}{2\Lambda_c} + (1 - \alpha)\r \log\l(2L - y)^2 - z^2 \r\,,
\eea
where $\tilde y_0 = 2l (B - \zeta \Delta \mu)/(B + k l)$.  For $k >  - \zeta \Delta \mu c_0 $, $\tilde y_0$ is greater than $2l$.
We also define the marginal distribution of the inclusion length $y$ as, 
\be \label{eq:marginal_length}
P(y) \equiv \int_{-(2L-y)}^{2L-y} dz P(y,z),
\ee
and the marginal distribution of the COM $z$
as 
\be \label{eq:marginal_COM}
P(z) \equiv \int_{0}^{2L-z} dy P(y,z).
\ee

\begin{figure}
   \centering
   \includegraphics[width=0.9\linewidth]{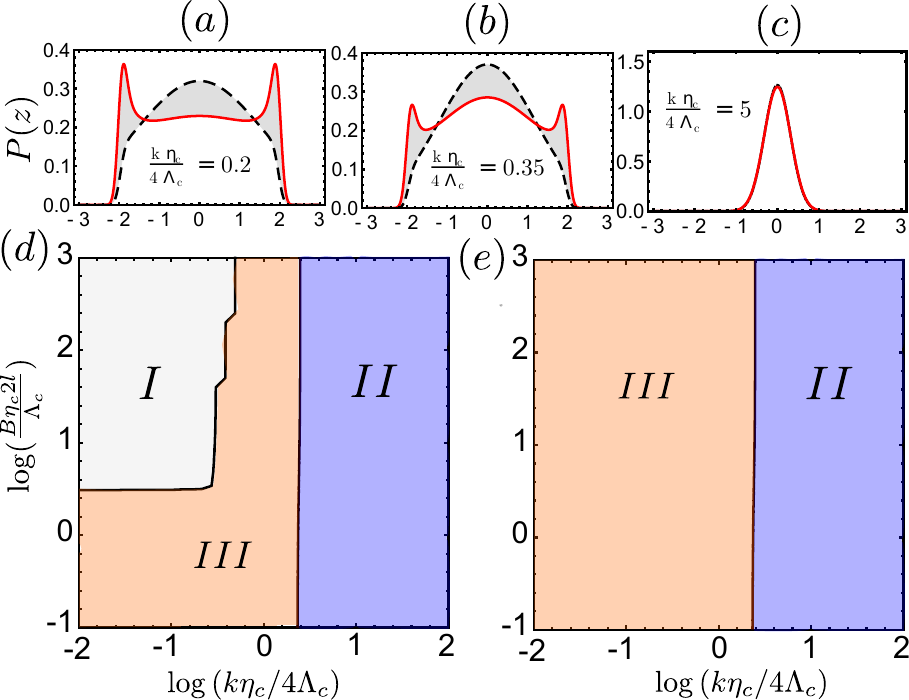}
   \caption{Marginal distribution  of the center of mass (COM, $z$), and phase plot of most likely COM position,  as a function of the dimensionless elasticity of the inclusion ($b=2l B \eta_c/\Lambda_c$) and the strength of the confinement ($k \eta_c/4\Lambda_c$), for active, and thermal fluctuations and parameter values $2L = 3, 2l = 1$, and $\zeta\Delta \mu c_0 \eta_c/\Lambda_c = -1$. 
(a--c)  Marginal distribution $P(z)$ of the COM, for active (solid red lines), and thermal (dotted black lines), for $2l B \eta_c/\Lambda_c = 50$ in the de-centered phase I (a), in the boundary phase (b) and in the centered phase II (c).  (d) Phase plot showing the most likely COM position for active fluctuations in the $(B,k)$ parameter plane. Different phases  correspond to  the inclusion being de-centred and extended ($\la y \ra> 2l$) (I),  centered and compressed ($y <2l$)(II), and centered and extended (III). Contrast this with the (e) phase plot of the most likely COM position for thermal fluctuations, which has only two phases, where the inclusion is, centered and compressed ($y <2l$)(II), and centered and extended (III).  }
	\label{fig:passive_px}
   \end{figure}

With these definitions, we find that, in contrast to the rigid case, the COM distribution for an elastic inclusion cannot be uniform.
We define the {\it centered phase} when the marginal distribution $P(z)$ at the center is highest over the region, else if the probability is maximum off-center, we define as the {\it de-centered phase}.
In the active case $\alpha=1/2$, the probability $P(z)$ is decentered (resp. centered) for low (resp. high) values of $k$, as shown in  Fig.\,\ref{fig:passive_px}(a)-(c) (solid red curve), contrast this with the thermal case (dotted black curve) where it is centered for all values of $k$. 
We find that for active fluctuations, the distribution of the COM position is either peaked on the boundary -- de-centered phase (I) -- or peaked at the center (centered phase), with inclusion being compressed (II), or extended (III). These transitions in the  phase diagram as a function of both the inclusion rigidity $B$ and the confinement strength $k$ are shown in Fig.\,\ref{fig:passive_px}(d). In contrast, as shown in Fig.\,\ref{fig:passive_px}(e), for thermal fluctuations there is only centered phase, with inclusion compressed (II),  or extended (III). 

In the next two paragraphs, we outline the main differences between thermal and active fluctuations; for simplicity, we first set to zero the boundary flow ($v_0 = 0$), the confining potential ($k = 0$), and the mean activity ($y_0 = 2l$). 
 
In the thermal noise case ($\alpha = 1$), Eq.\,\ref{eq:steady_potential} leads 
\be \label{eq:distribution_elastic}
P(y,z) = N \exp{\l -\f{B}{4l T}(y - y_0)^2 \r},
\ee
for all $-L + y/2\leq  z \leq L - y/2$, where $N$ is a normalization factor and $\Lambda_c/\eta_c = T$. As expected, Eq.\,\ref{eq:distribution_elastic} corresponds to a Boltzmann distribution with an elastic energy.
The marginal distribution for inclusion length is obtained by integrating Eq.\,\ref{eq:distribution_elastic} over the coordinate $z$
\bea \label{eq:marginal_distribution_elastic}
 P(y) = N (2L - y) \exp{\l -\f{B}{4l T}(y - y_0)^2 \r}, 
\eea
which is a Boltzmann distribution although not obvious at first sight.
 Indeed, the prefactor $(2L - y)$ in  Eq.\,\ref{eq:marginal_distribution_elastic} is due to the presence of the confinement, and the no-crossing condition on the inclusion boundary, which limits the $z$ integral range to $2L - y$. 

From Eqs.\,\ref{eq:distribution_elastic}, we estimate the moments of the inclusion length and position. For a large normalized elasticity $b$, the average and variance of the inclusion length converge to $\la y \ra = y_0 - (2l)^2/(2L - y_0)b$ and to $\mathrm{Var}[y] = \la y^2\ra - \la y\ra^2  = (2l)^2/b $, respectively, while the variance of COM position converges to $\la x^2\ra = \l (L - y_0/2)^2/3  + (2l)^2/b \r$. These limits match 
those expected for a rigid inclusion of length $y_0$. For small $b$, the asymptotic value of the averaged and variance of the inclusion length are $\la y \ra = 2L/3$ and $\mathrm{Var}[y] = 2L^2/9$, respectively, while the COM variance reads $\la x^2\ra = 2L^2/3$.

In the case of active fluctuations $\alpha = 1/2$, we find that the joint probability distribution reads
\be
 \nn P(y,z) = \frac{N}{\sqrt{(2L-y)^2 - z^2}} \exp \l - \f{B}{4l T} (y- y_0)^2 \r.
\ee
Integrating over $z$, we find that the marginal distribution on the inclusion length reads:
 \bea \label{eq:marginal_distribution_elastic_active}
P(y) &=& N \exp \l - \f{B}{4l T}(y-  y_0)^2 \r,
 \eea 
where $N$ is a normalization factor. Note that Eq.\,\ref{eq:marginal_distribution_elastic_active} differs by a prefactor $2L - y$ from the thermal case expression, Eq.\,\ref{eq:marginal_distribution_elastic}. 

From Eq.\,\ref{eq:marginal_distribution_elastic_active}, we estimate the moments of the inclusion length and position. For a large normalized elastic modulus $b$, we find that the mean and variance of the inclusion  length are, to leading order in ($1/b$), $\la y\ra = y_0 $ and $\la y^2\ra - \la y\ra^2  =  (2l)^2/b$, respectively. Notice that neither the mean nor the variance depends on the confinement length $L$. To leading order, the variance in the COM position read $\la x^2\ra = \l (L - y_0/2)^2/2 + (2l)^2/(2b) \r$, as expected by comparison with the the rigid inclusion case.  For small $b$ the asymptotic value of average length is $L$, variance is $L^2/3$, and the COM variance is $2L^2/3$.
 
\begin{figure}[t!]
 \centering
 \includegraphics[width=90mm]{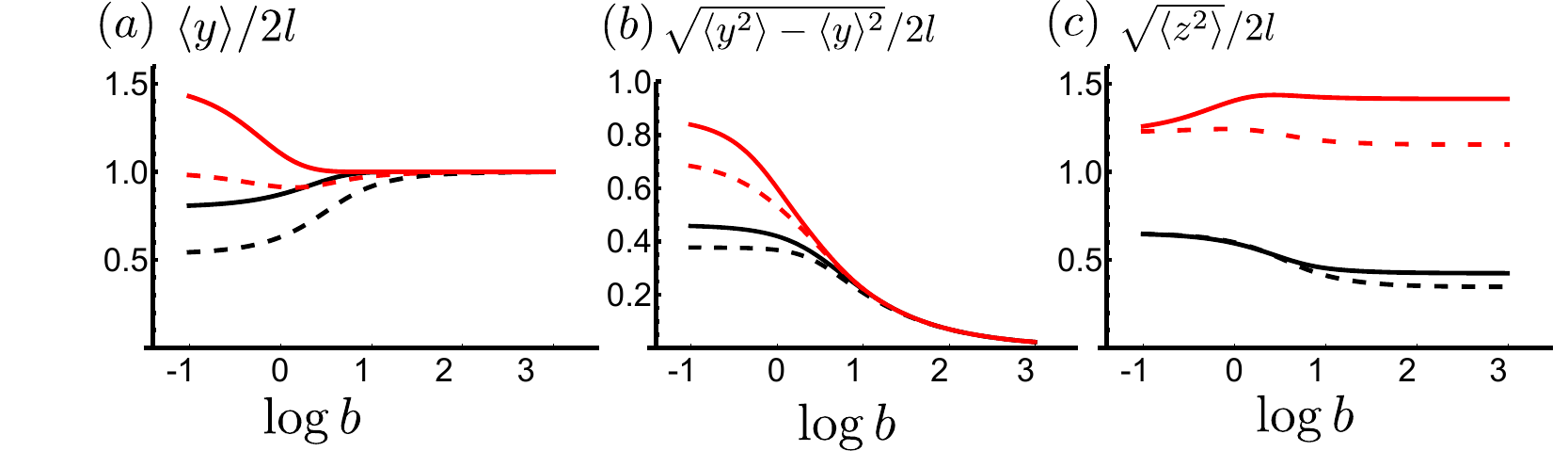}
 \caption{Moments as a function of the normalized elastic modulus $b$: (a) mean and (b) standard deviation of the inclusion length $y$, and (c) variance of the inclusion COM position $z$. Plotted for two sets of parameters $L/l =  1.6$ (black curve), and $L/l = 3$ (red curve), where we have fixed $y_0 = 2l = 1$, for both thermal fluctuation (dotted line), and active fluctuation (solid line). 
 Thermal fluctuations exhibit reduced moments compared to the active, for both sets of parameters. For large $b$, we recover the rigid inclusion moments (see Sec.\,\ref{sec:rigid}), while for small $b$ the moments converge to the calculated asymptotic limit (see Sec.\,\ref{sec:elastic}). 
      }
     	\label{fig:passive_modes}
     	  \end{figure}

We display the behavior of the moments in the inclusion length and COM position in Fig.\,\ref{fig:passive_modes} (dotted lines for thermal noise, solid lines for active noise).  In all cases the value for the thermal fluctuations is smaller than that for active fluctuations of same strength.
The differences in the mean and variance of the inclusion length is quite significant for small values of $b$ and decreases with increase in $b$; contrast this with the variance of COM for which the differences grow with increasing value of $b$.  
Note that unlike in Fig.\,\ref{fig:passive_px}(a),
the variation of $\la y \ra$ in Fig.\,\ref{fig:passive_modes}(a) is a fluctuation effect.  
For $L/l  < 2$ we observe a rather surprising behavior - the mean length of the inclusion for large fluctuation (small stiffness) is less than for small fluctuations (large stiffness) - as opposed to the case when 
$L/l > 2$ where the average length increases with increase in fluctuation. This behavior can be understood in terms of the asymptotic values mentioned, for large $b$ its  the free length $2l$, for small $b$ its governed by the confinement size, and is equal to $L$. 

\section{Kelvin-Voigt Inclusion} \label{sec:viscoelastic}

Lastly, we consider the case when the inclusion is viscoelastic of a Kelvin-Voigt type (i.e., it behaves elastically at the longest time scales). This situation is closer to a realistic description of the cell nucleus embedded in the active cytoplasm, since, as reported in  \cite{Versaevel2012,Makhija2016,Ramdas2015},  the noise on the nucleus is dominated by
active cytoskeletal processes.
As in the previous section, the unloaded length of the inclusion is denoted $2l$, the displacement from the unloaded configuration is $u$ and the elastic modulus is $B$; thus the elastic stress is equal to $B \ddx u$. In addition, we include the dissipative contribution ($\eta_{I} \ddx \dot u$) into the stress equation
\be
\sigma^{I}= B \ddx u + \eta_{I} \ddx \dot u,
\label{eq:siginclusion}
\ee
where $\eta_{I}$ is the internal viscosity of the inclusion.
Following the analysis in the previous sections, force balance within the inclusion implies that $\sigma^{I}=\sigma^{I}(t)$; integrating Eq.\,\ref{eq:siginclusion} over the reference length $x -l$ to $x + l$ leads to the relation, 
$$2l \sigma^{I}(t) = B( y - 2l) + \eta_{I} \dot y\,.$$

From the force balance condition on the two inclusion edges, we obtain the following multivariate Langevin equation on the variables ${\bf X } =  \begin{bmatrix} y & z \end{bmatrix}^T$:
\be \label{eq:langevin_viscoelastic}
\dot{{\bf X}} =  {\bf F} + {\bf G} \cdot {\bf \theta},
\ee
\begin{widetext}
where the drift force is
\be
\nn {\bf F} = \frac{1}{\gamma(y)}\begin{bmatrix} -(2L-y) \frac{B}{2l}(y - y_0) -\eta_c v_0 + f(x_1)(L+ x_1) + f(x_2)( x_2 - L) \\  z \l \frac{B}{2l}(y - y_0) - \frac{\eta_{I}}{2l }  v_0 \r + \frac{(\gamma(y) + \frac{\eta_{I}}{2l}z) f(x_2) (x_2 - L )}{\eta_c} +  \frac{(-\gamma(y) + \frac{\eta_{I}}{2l}z) f(x_1)(L + x_1)}{\eta_c}
\end{bmatrix},
\ee
\end{widetext}
the noise amplitude matrix is
\be
\nn {\bf G} = \begin{bmatrix}
 \sqrt{2\Lambda_c(L + x_1)}/\gamma(y) &    \sqrt{2\Lambda_c(L - x_2)}/\gamma(y) \\ 
 \frac{(-1 + \frac{\eta_{I} z}{2 l \gamma(y)})\sqrt{2\Lambda_c(L + x_1)}}{\eta} & \frac{(1 + \frac{\eta_{I} z}{2l \gamma(y)})\sqrt{2\Lambda_c(L - x_2)}}{\eta}  \end{bmatrix}, 
\ee
${\bf \theta}  = \begin{bmatrix} \theta_1 & \theta_2  \end{bmatrix}^T$ is Gaussian white noise of unit variance; 
$\gamma(y) \equiv \eta_c+ \eta_{I} (2L-y)/(2l)$ is a friction that depends on the inclusion length $y$, and $y_0 \equiv  2l \l 1 - (\zeta \Delta \mu c_0)/B \r$ is  the activity-renormalized rest length. 

For a general confining  force, the Fokker-Planck equation associated with Eq.\ref{eq:langevin_viscoelastic} does not satisfy the potential condition (Eq.\,\ref{eq:potentialcondition}), which implies the existence of a non-zero probability current $J_i$, even at steady state. 
This can be seen from the fact that keeping the dissipation term in the inclusion without the corresponding fluctuation source, violates FDT, and makes it essentially a two-temperature problem, with inclusion temperature set to zero. 

To analyze the dynamics when fluctuations are thermal, i.e., when both the inclusion and the surrounding fluid are at equal temperature $T$, we set $\Lambda_c = \eta_c k_B T$, and inclusion stress is given by, 
\be
\sigma^{I}= B \ddx u + \eta_{I} \ddx \dot u + \sqrt{2 \eta_{I} k_B T} \, \theta,
\label{eq:thermalinclusion}
\ee
where $\theta$ is unit variance Gaussian white noise. The steady state probability distribution obtained using the inclusion stress in Eq.\ref{eq:thermalinclusion}, is exactly the same as that obtained for the elastic inclusion with thermal fluctuation (Eq.\,\ref{eq:distribution_elastic}) in Section \ref{sec:elastic}.

In the absence of any kind of external potential in Eq.\,\ref{eq:langevin_viscoelastic}, the potential condition is satisfied, and the steady state is $P(y,z) = N e^{- \phi}$, with
\bea
\nn \phi &=&\f{b}{r} \l\f{y}{2l} \l\f{\alpha }{b}-d-1 \r+\f{1}{2} \l\f{y}{2l} \r^2 (d+r+1)-\f{1}{3} \l \f{y}{2l}\r^3\r  \\
\nn   &+& \l (1 - \alpha) - \f{\eta^2_c v_0}{2\Lambda_c} \r \log ((2 L -y)^2-z^2) \\
&+&  2(\alpha -1) \log \gamma(y),
\eea

where $b = (2l B \eta_c)/\Lambda_c$ is the ratio of two stresses, the inclusion elastic stress  $2l B$ and the fluctuating stress $\Lambda_c/\eta_c$,
 $r = \eta_c/\eta_I$, the ratio of the two viscosities, and $d = L/l$ the ratio of size of confinement to the inclusion.

Note that in the limit $\alpha = 1$, the effective potential  $\phi$ is cubic, and not harmonic. This is because, as discussed above, for $\alpha = 1$ this does not reduce to an isothermal description. It corresponds to a two-temperature problem, where the inclusion is connected to a bath at temperature zero, and surrounding fluid to a bath at temperature $\Lambda_c/k_B\eta_c$. This makes it inherently non-equilibrium, with an equilibrium description, in terms of an effective potential which has a very different form the actual potential of the system. 
In the following, we focus on the effect of active fluctuations, for which we take $\alpha = 1/2$, while we fix the other parameters at 
$v_0 = 0$, $\zeta = 0$, $d=L/l=4$ and $y_0 = 2l=1$.

\begin{figure}
   \centering
   \includegraphics[width=\linewidth]{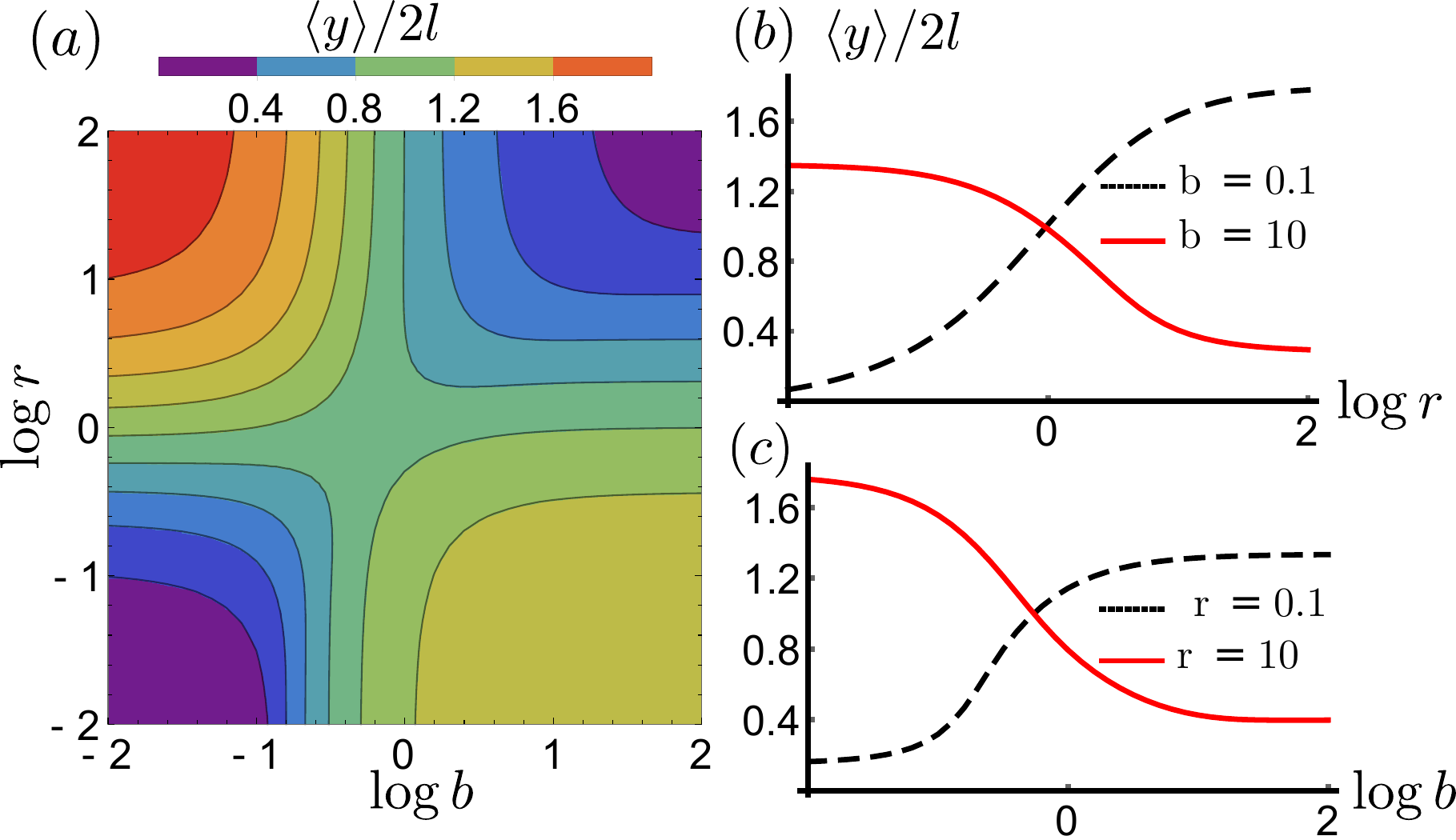}
   \caption{(a) Contour plot of the mean inclusion size $\la y \ra$ as function of the stress ratio $b$ and the viscosity ratio $r$, keeping the other parameters fixed (see text). (b,c) Mean inclusion size $\la y \ra$ versus $r$ and $b$, respectively.   
	    } 
	\label{fig:viscousavg}
   \end{figure}

Fig.\,\ref{fig:viscousavg}(a) shows a contour plot of the average inclusion length as function of stress ratio $b$, and viscosity ratio $r$.
In Fig.\,\ref{fig:viscousavg}(b), we take a cut across the contour plot, to show the average length as a function of viscosity ratio for two different values of $b$, the stress ratio. We find that for softer inclusions ($b=0.1$), the effective inclusion size is negligibly small when the inclusion viscosity is large (compared to the cytoplasm), and increases to beyond its unloaded length $2l$ when the inclusion viscosity is small. On the other hand for stiffer inclusions ($b=10$), the effective inclusion size is larger than $2l$ when the inclusion viscosity is large, and shrinks to below $2l$ when the inclusion viscosity is small. 
Similarly, Fig.\,\ref{fig:viscousavg}(c) is plotted with respect to the stress ration $b$. Unexpectedly, the inclusion shrinks either when the inclusion stiffness is increased or when the viscosity is decreased. 

\begin{figure}
   \centering
   \includegraphics[width=\linewidth]{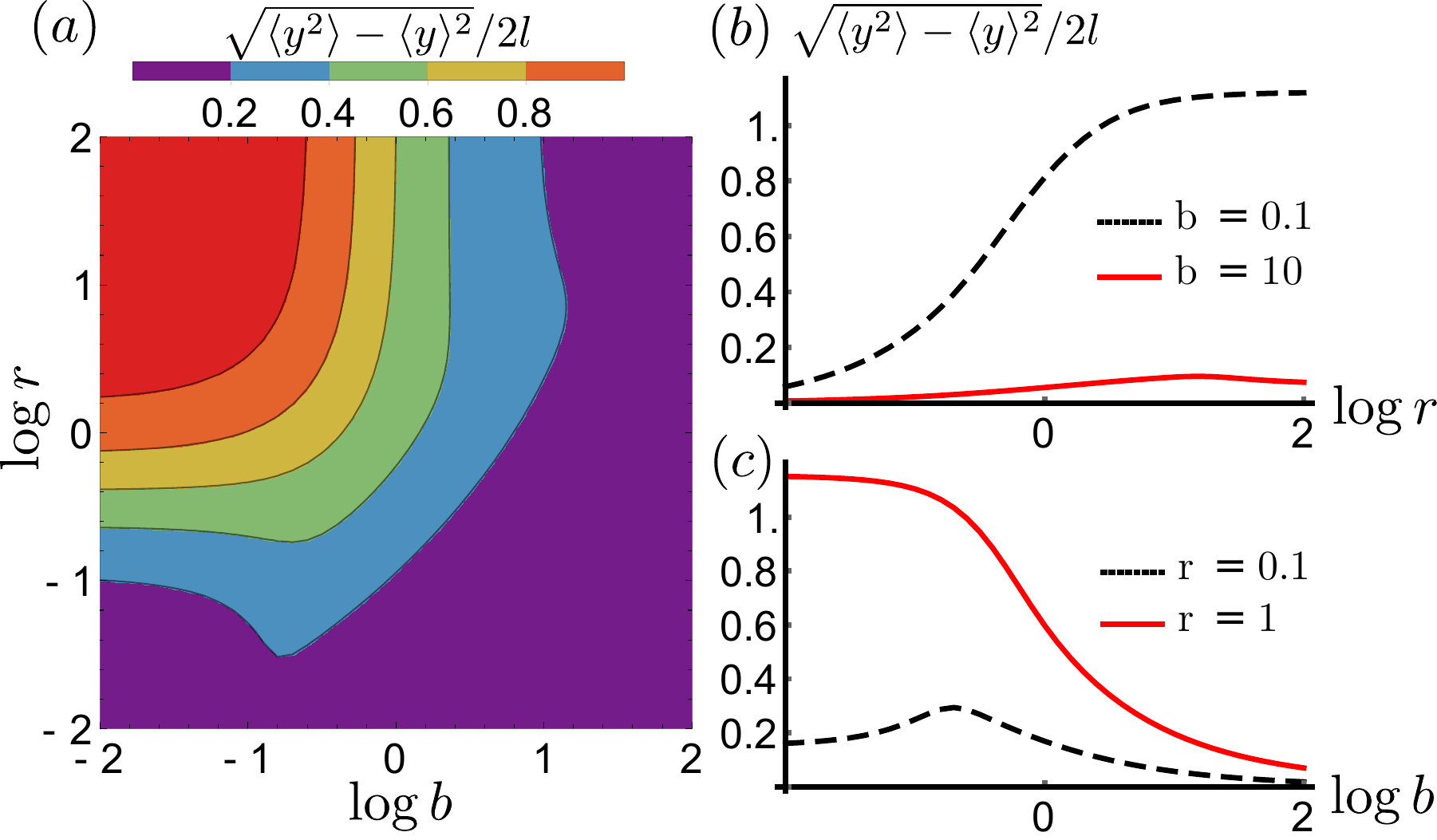}
   \caption{(a) Contour plot of the standard deviation of inclusion size $\sqrt{\la y^2 \ra - \la y \ra^2}$ as function of the stress ratio $b$ and the viscosity ratio $r$, all other parameters fixed. (b,c) Standard deviation of inclusion size $y$ versus $r$ and $b$, respectively.  
	    } 
	\label{fig:viscousvar}
   \end{figure}

\begin{figure}
   \centering
   \includegraphics[width=\linewidth]{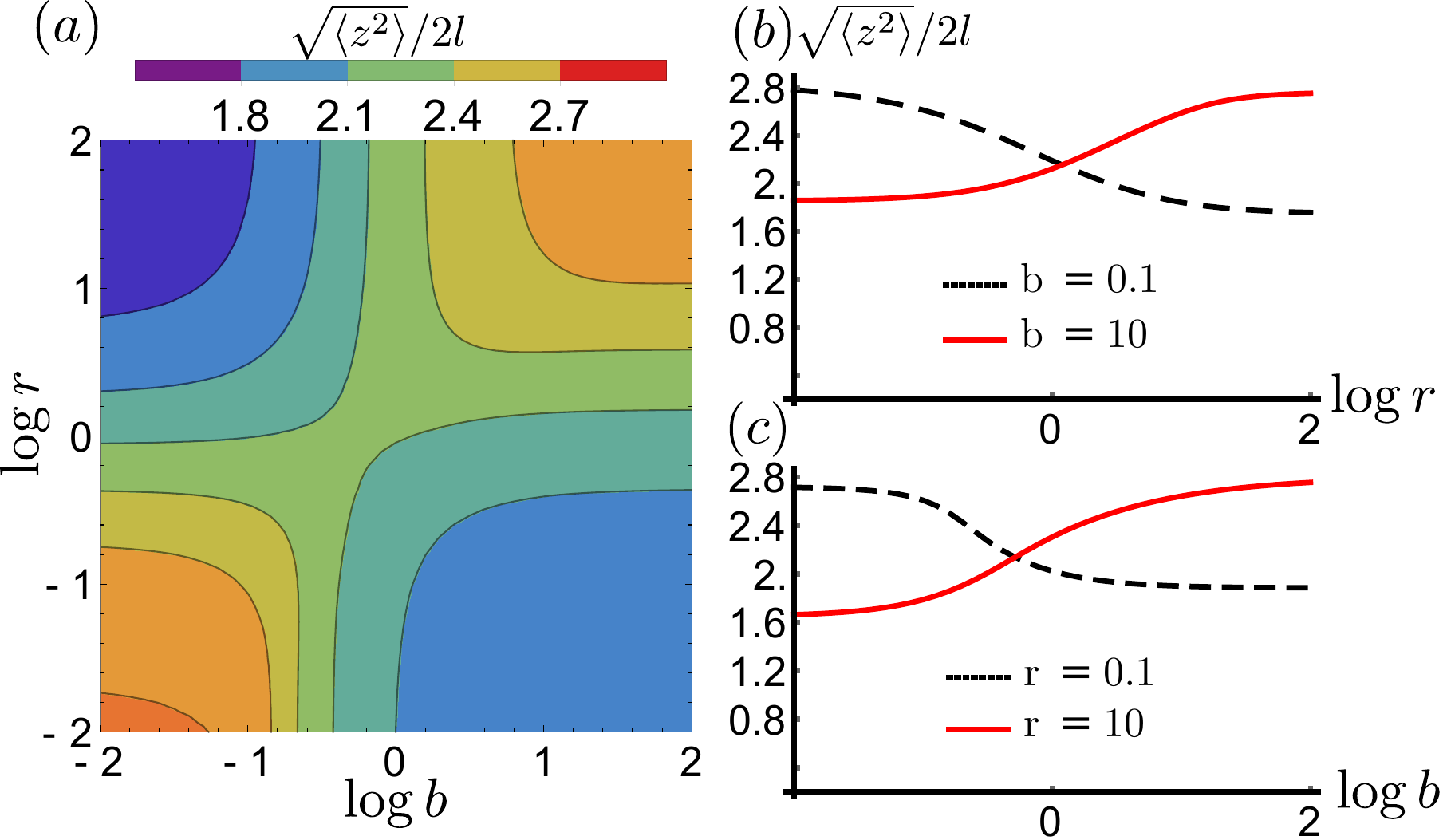}
   \caption{(a) Contour plot of the standard deviation of centre of mass position of the inclusion $\sqrt{\la z^2 \ra}$ as function of the stress ratio $b$ and the viscosity ratio $r$, all other parameters fixed. (b,c) Mean centre of mass position $\sqrt{\la z^2 \ra}$ versus $r$ and $b$, respectively.  
	    } 
	\label{fig:viscousvarz}
   \end{figure}

We display in Figs.\,\ref{fig:viscousvar} and \ref{fig:viscousvarz} how the amplitude of fluctuations - of both the width and COM position - vary as a function of $b$ and $r$.
As expected, a stiffer inclusion generally correspond to a lower amplitude in the fluctuations of the width; less intuitive is the observation that a  stiffer inclusion leads to an increase of fluctuations in COM.

\section{Force on the confining walls situated at the cell boundary}\label{sec:force}
We now determine the force induced by the fluctuating inclusion at the confining walls situated at the cell boundary. This force depends on the specific nature of the fluctuations, and is a non-equilibrium Casimir force  \cite{Aminov2015,Brito2007, Kirkpatrick2013}.

Not surprisingly, the form of the force depends on the physical nature of the inclusion. In the case of a rigid inclusion and in the absence of a confining potential, the local stress at the boundary reads  $\sigma = - \zeta \Delta \mu c_0 + \l   \sqrt{2\tilde L} \theta \r/(2\tilde L) $, whose mean is $\la \sigma \ra  = - \zeta \Delta \mu c_0$.  There is no fluctuation contribution to the average boundary force, irrespective of whether the noise is  thermal \cite{Monahan2015} or active.

In the elastic inclusion case, the average stress  on the confining walls is given by $ \la \sigma \ra = B (\la y\ra - 2l)/(2l)$. 
As discussed in Section \ref{sec:elastic}, the average inclusion size $\la y \ra$ depends on the strength and nature of the noise (whether thermal or active). 

\begin{figure}
   \centering
   \includegraphics[width=0.9\linewidth]{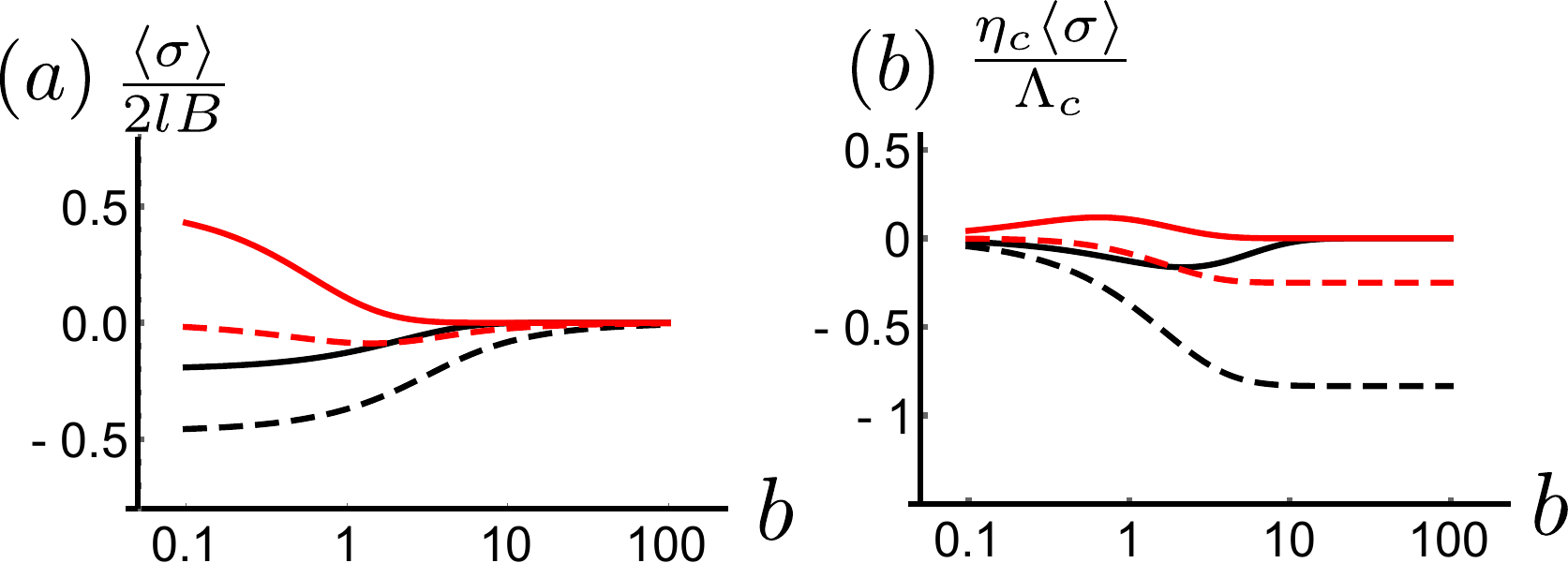}
   \caption{Average stress for an elastic inclusion embedded in a viscous fluid subject to thermal (dotted line) and active (solid line)
   noise, for $2l = 1$, $2L = 2$ (black), and $2L = 3$ (red) as function of $b$, the ratio of the elastic to the fluctuating stress. In (a) the stress is expressed in units of the elastic stress $\la \sigma \ra/2l B$ and plotted as a function of $\eta_c/\Lambda_c$, for fixed $B$, while in (b) the stress is in units of the fluctuation stress $\la \sigma \ra \eta_c/\Lambda_c$ and plotted as a  function of $B$, for fixed $\eta_c/\Lambda_c$. 
	    } 
	\label{fig:stress}
   \end{figure}

In Fig.\,\ref{fig:stress}(a) we assume a fixed stiffness; we represent the average stress as a function of the inverse amplitude of fluctuation.
When the amplitude of fluctuation $\Lambda_c/\eta_c$ is weak compared to the elastic stress (large $b$), the fluctuation contribution to the force on the walls converges to zero -- this corresponds to the rigid inclusion result. For large fluctuations,  the sign of the force depends on $L/l$, the ratio of size of confinement to size of inclusion. 
For comparable size (black curve) it is attractive, and for an inclusion much smaller than confinement scale (red curve), it is  repulsive (positive value of stress).

The average stress is a decreasing function of the elastic modulus $B$ when the confinement size is large compared to the inclusion width (as seen in Fig.\,\ref{fig:stress}(b), red curves). However, the average stress can become an increasing function of $B$ for larger inclusion width. As expected from the rigid inclusion case, the force on the walls vanishes when the elastic strength $2l B$ becomes small compared to the fluctuation strength $\Lambda_c/\eta_c$\,. For  large stiffness $B$, the asymptotic value of the averaged stress depends on whether the noise is thermal or active : for thermal noise, the averaged stress becomes attractive, with an amplitude that is controlled by the ratio $L/l$; for active noise, the averaged stress vanishes. 
The presence of an effective force between the walls, even for thermal fluctuations, shows that inhomogeneity in form of inclusions has nontrivial implications.

%
%
%
%
%
%
%
%
\begin{widetext}

\label{tab:steadystate}\begin {table}[h!]
{
\begin{center}
\caption {Steady state distribution $P(y,z)$ of $y$ (inclusion length) and $z$ ($2 \times$center-of-mass), in absence of external potential ($V = 0$), and no boundary flux ($v_0 = 0$),
up to a normalization factor. See Table I for definition of parameters.} \label{tab:summary} 
\begin{tabular}{|c|c|c|}
  \hline
  Inclusion & Thermal fluctuation & Active fluctuation \\
  \hline
  \rule{0pt}{3ex}
  Rigid & $P(z)=1$ & $P(z)=\f{1}{\sqrt{(2L -2l)^2 - z^2}}$ \\
    \hline
  \rule{0pt}{5ex}
  Elastic & $P(y,z)=e^{- \f{B \eta_c}{2l \Lambda_c}\l y - 2l \r^2}$ & $P(y,z)=\f{e^{- \f{B \eta_c}{2l \Lambda_c}\l y - 2l \r^2}}{\sqrt{(2L -y)^2 - z^2}}$ \\
      \hline
    \rule{0pt}{5ex}
  Kelvin-Voigt & $P(y,z)=e^{- \f{B \eta_c}{2l \Lambda_c}\l y - 2l \r^2}$ & $P(y,z)=\f{\l(2L- y) r + 2l \r\exp\l - \l \f{b}{r} \l\f{y}{2l} \l\f{1 }{2 b}-d-1 \r+\f{1}{2} \l\f{y}{2l} \r^2 (d+r+1)-\f{1}{3} \l \f{y}{2l}\r^3\r\r \r}{\sqrt{(2L -y)^2 - z^2}}$ \\
  \hline
\end{tabular}
\end{center}
}
\end {table}

\end{widetext}

\section{Conclusion}

Starting from 1D fluctuating active hydrodynamics we derive a Langevin dynamics description of a point inclusion embedded in an active gel. 
We find that the stochastic dynamics of the inclusion is given by a generalized underdamped Langevin equation with multiplicative noise. 
We obtain the Overdamped limit by adiabatically eliminating the three fast time scales of the problem - inertial relaxation time ($\tau_m$), maxwell time ($\tau_v$), and noise correlation time ($\tau_A$).   This gives us the right choice of stochastic calculus associated with the Overdamped Langevin equation. We show that the order in which the times scales are integrated out is important, instead of $\tau_v$, the crucial time scale is $\sqrt{\tau_m \tau_v}$. 

We then extend this analysis to three different extended inclusion types, rigid, elastic, and viscoelastic (Kelvin-Voigt), where we take the embedding medium to be fluid. 
The stochastic dynamics of inclusion length and center of mass is given by a set of coupled Langevin equation with multiplicative noise.
For all these cases we can analytically obtain the steady state probability distribution. Table I lists the steady state
distribution of the center of mass and length of the inclusion for these three inclusion types, for both thermal and active fluctuations.

 In the simplest description of the inclusion as a rigid object, we obtain analytical expression for the steady state distribution, which reveal the existence of a fluctuation-induced effective potential.
 This attractive force, which originates from the non-equilibrium nature the noise, is reminiscent of Casimir forces in non-equilibrium systems \cite{Brito2007, Kirkpatrick2013, Aminov2015}.

Considering the inclusion as an elastic element introduces an additional degree of freedom along with centre of mass, namely the extension or the inclusion length. The effective attraction between the inclusion and the confining walls persists, with the noise induced force dependent on the length of the inclusion.  We find that the average length of the elastic inclusion depends non-trivially on the size of confinement and strength of the fluctuation.

 Unlike the rigid and elastic description, a Kelvin-Voigt viscoelastic description of the inclusion, in general cannot be mapped to an equilibrium system with an effective potentials. 
Nevertheless, it is possible to have an effective equilibrium description in the absence of external confining forces.  We find that the viscosity of the inclusion is a crucial parameter, the mean and variances of the inclusion length strongly depend on the viscosity ratio of the inclusion and the outside fluid, along with its stiffness and stress fluctuation.

Since the observed phenomena are due to the multiplicative and active nature of the noise, we believe these noise induced interactions should  occur in a variety of biological and in-vitro contexts. We list a few
here:
(i) the positioning and shape fluctuations of the nucleus (or other localized organelles) within a cell, 
(ii) the dynamics of large colloidal particles embedded in an active medium, (iii) the positioning and dynamics of nuclei in multi-nucleated cells,
 (iv) the positioning and segregation of chromosomes within a nucleus and  (v) the fluctuations of a cell within a 
developing tissue. 
For instance, chromosomes within the cell nucleus,  can be thought of as 
soft ellipsoidal inclusions embedded within an active nucleoplasm \cite{Wang2017,Uhler2017}. A theoretical study of the relative positioning, orientation and overlap of these soft ellipsoids would require a generalization of the ideas developed in this paper to 3-dimensions. On the other hand, in the tissue context, one could study the dynamics of tissue vertex or tissue junctions as active inclusions in an active tissue medium.

Our study of the active Casimir-like forces  at the cell boundary, arising from nuclear stiffness and fluctuations might be relevant to rigidity sensing by focal adhesions \cite{Cao2015}.

We are currently extending our study to the case when the (nuclear) inclusion is subject to its own source of active fluctuation, in addition to the noise coming from the surrounding active fluid.
We are also studying the collective dynamics of  colloids \cite{Happel, Doi1988} embedded in an active fluid in higher dimensions, where the hydrodynamic interactions lead to multiplicative noise.

\section{Acknowledgement}
We thank S. Ramaswamy, F. Julicher, K. Vijaykumar, and R. Morris for clarifying discussions, as well as A. Rautu and K. Husain for help in the manuscript.  AS thanks MBI-Singapore and J.-F.R. thanks NCBS-Bangalore for hospitality.

\appendix

\section{Constant total myosin number}\label{ap:constant_c}
If  the myosin turnover is slow compared to the time scales of interest, the total number of myosin in the two segments L(R) can be taken to be constant $c_{L(R)}$. With this, we obtain from Eqs.\,\ref{eq:inclusiondynam_1},\,\ref{eq:inclusiondynam_2},
\bea
\eta_c \dx_1 &=& \eta_c v_L + (x_1 +  L) \sigma^L(t) + \zeta \Delta \mu c_{L} -  \int^{x_1}_{-L} \fc \,dx, \nonumber \\
\nn \eta_c \dx_2 &=& -\eta_c v_R + (x_2 - L) \sigma^R(t) - \zeta \Delta \mu c_{R} +  \int^{L}_{x_2} \fc \,dx.
\eea
Comparing this with Eqns.\,\ref{eq:x1},\,\ref{eq:x2}  we see that, the effect of the constant myosin number is same as having a boundary flow $v_0=\zeta \Delta \mu c_{L}/\eta_c$. 
In this paper, we have treated the mean contractile stress $\zeta \Delta \mu$ and noise amplitude $\Lambda$ as independent. In reality, they have a common origin in the actomyosin remodeling
dynamics and $\Lambda$ depends on $\zeta\Delta \mu$ \cite{Basu2008}.
  
If $\Lambda= \Lambda_0 c$, the actomyosin density, and we assume slow turnover of actomyosin, then the noise $\theta$ is additive and the equations reduce to
\bea
\eta_c \dx_1 &=& \eta_c v_L + (x_1 +  L) \sigma^L(t) + \zeta \Delta \mu c_{L} -  \sqrt{2\Lambda_0 c_L}\theta_1, \nonumber \\
\nn \eta_c \dx_2 &=& -\eta_c v_R + (x_2 - L) \sigma^R(t) - \zeta \Delta \mu c_{R} +  \sqrt{2\Lambda_0 c_R}\theta_2,
\eea
where $\theta_{1(2)}$ are unit variance Gaussian white noises.

\section{Langevin to Fokker-Planck}\label{ap:langevin_to_fp}
It is well known that  the overdamped Langevin equation with multiplicative noise is ill-defined \cite{Kampen1981,Gardiner},
here we follow the derivation in \cite{Lau2007}.
The multi-variate overdamped Langevin equation with multiplicative noise is, 
\bea
\dot x_i = f_i(x) + g_{ij}(x)\, \theta_j,
\label{eq:multiplicative}
\eea
where $i$ varies from $1$ to $N$, $j$  varies from $1$ to $M$ and it is understood that repeated indices are summed over. $\theta$ is an uncorrelated zero mean, Gaussian white noise, and $\la \theta_i(t) \theta_j(t') \ra = \delta_{ij} \delta(t - t')$. 
Integrating (\ref{eq:multiplicative}) over time interval $\Delta t$, we get, 
\bea
\nn \Delta x_i &=& \int_{t}^{t + \Delta t} dt f_i(x) + \int_{t}^{t + \Delta t} dt \, g_{ij}(x) \theta_j,
\eea
where $\Delta x_i$ is the displacement in time $\Delta t$. Unlike the integral of the deterministic force term $f(x)$, the integral of the fluctuation term $g_{ij}(x) \theta_j$ depends on the time point at which $g(x)$ is evaluated \cite{Gardiner, Lau2007}.  For every $i,j$, the integral of the stochastic term is approximated by, 
\be
\int_{t}^{t + \Delta t} dt \, g_{ij}(x) \theta_j = g_{ij}(x) \int_{t}^{t + \Delta t} dt  \theta_j .
\ee
In $g_{ij}(x)$,  $x$ can be either evaluated at time $t$ (Ito-convention), or at the mid-value $x = x(t)  + \Delta x /2$ (Stratonovich-convention), or at $x = x(t) + \Delta x$ (Hanggi-Klimontovich-convention)
\cite{Hanggi1982, Klimontovich1990}.

In general, any time point can be chosen \cite{Lau2007}. If we evaluate at time $ t + \alpha \Delta t$, we get,
\bea
\nn \Delta x_i &=&  f_i(x(t)) \Delta t +  g_{ij}(x(t) + \alpha \Delta x))\int_{t}^{t + \Delta t} dW_j.
\eea
Considering the Taylor expansion of $g_{ij}$ for small $\Delta x$ leads to
\bea
\nn \Delta x_i &=&  f_i(x(t)) \Delta t +  g_{ij}(x(t))\int_{t}^{t + \Delta t} dW_j \\
&+& \alpha \f{\p g_{ij}}{\p x_k} \Delta x_k\int_{t}^{t + \Delta t} dW_j,
\eea
Substituting $\Delta x_k$ back in the equation and keeping first two terms, we get, 
\bea
\nn \Delta x_i &=&  f_i(x(t)) \Delta t +  g_{ij}(x(t))\int_{t}^{t + \Delta t} dW_j, \\
 &+& \alpha \f{\p g_{ij}}{\p x_k} g_{kl}(x(t))\int_{t}^{t + \Delta t} dW_l \int_{t}^{t + \Delta t} dW_j.
\eea
From this, we obtain following equation for the mean and variance,
\bea
\la \Delta x_i \ra  &=&  f_i(x(t)) \Delta t  +  \alpha \f{\p g_{ij}}{\p x_k} g_{kj}(x(t)) \Delta t, \\
\la \Delta x_i \Delta x_j \ra &=& g_{ik} g_{jk} \Delta t.
\eea
The Fokker-Planck equation corresponding to this is \cite{Gardiner}, 
 \bea
\nn \p_t P &=& \f{\p}{\p x_i} \l - f_i(x) - \alpha \f{\p g_{ij}}{\p x_k} g_{kj} + \f{\p}{\p x_j}  g_{ik} g_{jk} \r P. \\
\label{eq:FP_alpha}
\eea
Thus we see that the Fokker-Planck equation depends on the choice of $\alpha$. If instead the noise is additive, the $\alpha$ dependent term in Eq.\ref{eq:FP_alpha} is identically zero, and
hence the convention does not matter.

\section{Overdamped Langevin equation from generalized Langevin dynamics}\label{ap:limit}

In the previous Appendix, we showed that the overdamped Langevin equations with multiplicative delta-function noise are ill defined - they result in different Fokker-Planck descriptions under different choices of stochastic calculus 
used to discretize the noise term. Hence, overdamped Langevin equations with multiplicative delta-function noise, must be provided with an interpretation of the noise, in order to be meaningful. 

An unambiguous approach is to start with the correct microscopic inertial dynamics for all the microscopic variables, and establish a separation of time scales. One then systematically integrates over 
the shorter time scales to arrive at the correct overdamped Langevin equations.
We start with a generalized underdamped Langevin dynamics for a particle position $x(t)$, with spatially varying damping and a noise $\theta$ that satisfies an Ornstein-Uhlenbeck process \cite{Gardiner},
\bea
\nn \dot x &=& v, \\
\nn m \dot v + \gamma(x) v &=& f(x)  + g(x) \frac{\theta}{\sqrt{\tau_n}}, \\
 \dot \theta &=& - \f{1}{\tau_n}\theta + \f{1}{\sqrt{\tau_n}} \vartheta,
\label{eq:langevin}
\eea
where  $\la \vartheta(t) \vartheta(t')\ra = 2 \Lambda \delta(t - t')$. The other relevant time scale is the inertial relaxation time $\tau_m = m/\gamma(x)$. 

To go from here to an overdamped Langevin equation with white noise, it is necessary to take the two limits : $\tau_m \to 0$ to get the overdamped dynamics, and
$\tau_n \to 0$, to get the white-noise limit. However, as discussed in \cite{Kupferman2004}, one might choose to take these limits in different  order, which result in different
interpretations of the overdamped equations. Since Langevin dynamics corresponding to thermal noise is constrained to obey the fluctuation-dissipation relation (FDR), it is convenient  to
treat the thermal and active noise cases separately.

\subsection{Thermal}

A necessary condition for the Langevin dynamics Eq. \ref{eq:langevin} to describe thermal noise 
is that 
 $\tau_n \rightarrow 0$, which leads to

\bea
 m \dot v + \gamma(x) v &=& f(x)  + g(x)\vartheta\, , \\
 \nn \dot x &=& v \, .
 \eea
The corresponding Fokker-Planck is,
\bea
\nn \p_t P &=&  - \p_x  v   P +  \f{1}{m}  \p_v \l \gamma(x) v -f(x)  +  \f{\Lambda}{m} g^2(x) \p_v \r P.  \\
\label{eq:thermal_FP}
\eea
To obtain the overdamped Langevin equation from the Fokker-Planck equation \ref{eq:thermal_FP}, we use the technique of
adiabatic elimination in the momentum variable \cite{Gardiner,Fox1986,Baule2009}.

Define the moments of $v$, $Q_k = \int dv \, v^k P $.  The equations for the moments of $v$ are,
\bea
\nn \f{\p}{\p t} Q_0 &=& - \f{\p}{\p x} Q_1,   \\
\nn \f{\p}{\p t} Q_1 &=& - \f{\p}{\p x} Q_2     - \f{\gamma(x)  }{m} Q_1  + \f{f(x)}{m} Q_0,\\
\nn \f{\p}{\p t} Q_2 &=& - \f{\p}{\p x} Q_3    - \f{2\gamma(x) }{m} Q_2 + \f{2 f(x)}{m} Q_1 + 2 \Lambda \f{g^2(x)}{m^2} Q_0,
\eea
and so on. Thus to solve for $Q_0$, we require knowledge of higher moments. However, we note that the $k^{th}$-moments $Q_k$ decay exponentially fast with a time scale proportional to 
$\tau_m$. Thus in the limit $\tau_m \to 0 $, we can assume the higher moments reach steady state, from which we obtain
\bea
 Q_2 &=&  - \f{m}{2\gamma(x)} \f{\p}{\p x} Q_3 +  \f{f(x)}{\gamma(x)} Q_1 + \Lambda \f{g^2(x)}{m \gamma(x)} Q_0, \\
 Q_1  &=&   - \f{m}{\gamma(x)}\f{\p}{\p x} Q_2   +  \f{f(x)}{\gamma(x)} Q_0.
\eea
Using these relation in the equation for $Q_0$, and ignoring terms of order $\tau_m$ and higher, we obtain 
\bea
\nn \f{\p}{\p t} Q_0 &=& \f{\p}{\p x} \l  -  \f{f(x)}{\gamma(x)} Q_0 +  \f{\Lambda}{\gamma(x)}\f{\p}{\p x} \l    \f{g^2(x)}{\gamma(x)} Q_0 \r  
 \r .
\eea

If the fluctuations are thermal then FDR holds, and $\gamma(x) =  g^2(x)$, and $\Lambda = k_B T$, where $T$ is temperature of the bath, and the above equation reduces to,
\bea
 \p_t Q_0 &=&  \p_x \l  -\f{f(x)}{\gamma(x)}   + \f{k_b T}{\gamma(x)}\f{\p}{\p x} \r Q_0.
\eea
The corresponding Langevin equation
\bea
\gamma(x) \dot x &=& f(x)  + g(x)\,\vartheta,
\label{eq:overdamped1}
\eea 
is interpreted in Hanggi-Klimontovich convention \cite{Hanggi1982, Klimontovich1990}.  
\subsection{Active}
On the other hand, by not setting $\tau_n$ to zero, we are necessarily describing a situation where the noise is athermal. This is consistent with an active noise,
where the microscopic variables (actomyosin remodeling and turnover) describing active noise are slower than the inertial relaxation time, $\tau_m \ll \tau_n$.

Taking $\tau_m \to 0$ in \ref{eq:langevin}, we get, 

\bea
  v &=& \f{f(x)}{\gamma(x)}  + \f{g(x)}{\gamma(x)}\frac{\theta}{\sqrt{\tau_n}}, \\
\nn  \dot \theta &=& - \f{1}{\tau_n}\theta + \f{1}{\sqrt{\tau_n}}\vartheta,
 \eea
The corresponding Fokker-Planck  is,
\bea
\nn \f{\p}{\p t} P &=&  - \f{\p}{\p x}  \l \f{f(x)}{\gamma(x)}  +  \f{1}{\sqrt{\tau_n}}\f{g(x)}{\gamma(x)} \theta  \r  P, \\
&  & +  \f{1}{\tau_n}\f{\p}{\p \theta} \l  \theta  + \Lambda \f{ \p}{\p \theta} \r P .
\label{eq:FP_active}
\eea
We define the moment of $\theta$, $Q_k = \int d\theta \, \theta^k P $. From \ref{eq:FP_active} we obtain 
\bea
\nn \f{\p}{\p t} Q_0 &=& - \f{\p}{\p x} \f{f(x)}{\gamma(x)} Q_0 - \f{1}{\sqrt{\tau_n}}\f{\p}{\p x} \f{g(x)}{\gamma(x)} Q_1, \\
\nn \f{\p}{\p t} Q_1 &=& - \f{\p}{\p x} \f{f(x)}{\gamma(x)} Q_1 - \f{1}{\sqrt{\tau_n}}\f{\p}{\p x} \f{g(x)}{\gamma(x)} Q_2 -    \f{1}{\tau_n} Q_1, \\
\nn \f{\p}{\p t} Q_2 &=& - \f{\p}{\p x} \f{f(x)}{\gamma(x)} Q_2 - \f{1}{\sqrt{\tau_n}}\f{\p}{\p x} \f{g(x)}{\gamma(x)} Q_3  -     \f{2}{\tau_n} Q_2 +  \f{2\Lambda}{\tau_n} Q_0,
\eea
and so  on. 
Following the arguments in the case of thermal noise, we obtain:  
in the limit $\tau_n  \to 0 $, 
\bea
  Q_1 &=&  - \sqrt{\tau_n}\f{\p}{\p x} \f{g(x)}{\gamma(x)} Q_2 -  \tau_n\f{\p}{\p x} \f{f(x)}{\gamma(x)} Q_1, \\
 Q_2 &=&  -\f{\sqrt{\tau_n }}{2}\f{\p}{\p x} \f{g(x)}{\gamma(x)} Q_3  - \f{\tau_n }{2}  \f{\p}{\p x} \f{f(x)}{\gamma(x)} Q_2   +    \Lambda Q_0, \qquad 
\eea
which leads to 
\bea
 \f{\p}{\p t} Q_0 &=& - \f{\p}{\p x} \f{f(x)}{\gamma(x)} Q_0 + \Lambda \f{\p}{\p x} \f{g(x)}{\gamma(x)} \f{\p}{\p x} \f{g(x)}{\gamma(x)}  Q_0. \quad
\eea
The corresponding Langevin equation
\bea
\gamma(x) \dot x &=& f(x)  + g(x)\,\vartheta,
\label{eq:overdamped}
\eea 
is interpreted in the Stratonovich convention\cite{Stratonovich, Gardiner} .

\section{Multiple noise sources}\label{ap:multiple_noise}
The Langevin equation driven by multiple noise sources labeled by $i =1, \ldots, N$, each of which are exponentially  correlated over different times scales, can be described by
the following  Ornstein-Uhlenbeck processes,
\bea
\dot x &=& f(x) + g_i(x) \theta_i ,\\
\tau_i \dot \theta_i &=& - \theta_i + \vartheta_i,
\eea
 where the  variance of Gaussian white noise $\vartheta_i $ is  $2\Lambda_i$.  
Since all the noise sources are independent, for observation times $t \gg \text{max} (\tau_i)$, we can adiabatically integrate out each time scale, which will lead to the  following Fokker-Planck equation, 
\bea
\p_t P = \p_x \l - f(x) + \f{1}{2} \sum_i g_i(x) \p_x g_i(x)\r P.
\eea
Since $g_i$ has originated from the hydrodynamic interactions, its form is the same for all noise sources thus
\bea
\p_t P = \p_x \l - f(x) +  \l \sum_i \Lambda_i\r g(x) \p_x g(x)\r P. \quad
\eea


\begin{thebibliography}{50}%
\makeatletter
\providecommand \@ifxundefined [1]{%
 \@ifx{#1\undefined}
}%
\providecommand \@ifnum [1]{%
 \ifnum #1\expandafter \@firstoftwo
 \else \expandafter \@secondoftwo
 \fi
}%
\providecommand \@ifx [1]{%
 \ifx #1\expandafter \@firstoftwo
 \else \expandafter \@secondoftwo
 \fi
}%
\providecommand \natexlab [1]{#1}%
\providecommand \enquote  [1]{``#1''}%
\providecommand \bibnamefont  [1]{#1}%
\providecommand \bibfnamefont [1]{#1}%
\providecommand \citenamefont [1]{#1}%
\providecommand \href@noop [0]{\@secondoftwo}%
\providecommand \href [0]{\begingroup \@sanitize@url \@href}%
\providecommand \@href[1]{\@@startlink{#1}\@@href}%
\providecommand \@@href[1]{\endgroup#1\@@endlink}%
\providecommand \@sanitize@url [0]{\catcode `\\12\catcode `\$12\catcode
  `\&12\catcode `\#12\catcode `\^12\catcode `\_12\catcode `\%12\relax}%
\providecommand \@@startlink[1]{}%
\providecommand \@@endlink[0]{}%
\providecommand \url  [0]{\begingroup\@sanitize@url \@url }%
\providecommand \@url [1]{\endgroup\@href {#1}{\urlprefix }}%
\providecommand \urlprefix  [0]{URL }%
\providecommand \Eprint [0]{\href }%
\providecommand \doibase [0]{http://dx.doi.org/}%
\providecommand \selectlanguage [0]{\@gobble}%
\providecommand \bibinfo  [0]{\@secondoftwo}%
\providecommand \bibfield  [0]{\@secondoftwo}%
\providecommand \translation [1]{[#1]}%
\providecommand \BibitemOpen [0]{}%
\providecommand \bibitemStop [0]{}%
\providecommand \bibitemNoStop [0]{.\EOS\space}%
\providecommand \EOS [0]{\spacefactor3000\relax}%
\providecommand \BibitemShut  [1]{\csname bibitem#1\endcsname}%
\let\auto@bib@innerbib\@empty
\bibitem [{\citenamefont {Marchetti}\ \emph {et~al.}(2013)\citenamefont
  {Marchetti}, \citenamefont {Joanny}, \citenamefont {Ramaswamy}, \citenamefont
  {Liverpool}, \citenamefont {Prost}, \citenamefont {Rao},\ and\ \citenamefont
  {Simha}}]{Activereview}%
  \BibitemOpen
  \bibfield  {author} {\bibinfo {author} {\bibfnamefont {M.~C.}\ \bibnamefont
  {Marchetti}}, \bibinfo {author} {\bibfnamefont {J.~F.}\ \bibnamefont
  {Joanny}}, \bibinfo {author} {\bibfnamefont {S.}~\bibnamefont {Ramaswamy}},
  \bibinfo {author} {\bibfnamefont {T.~B.}\ \bibnamefont {Liverpool}}, \bibinfo
  {author} {\bibfnamefont {J.}~\bibnamefont {Prost}}, \bibinfo {author}
  {\bibfnamefont {M.}~\bibnamefont {Rao}}, \ and\ \bibinfo {author}
  {\bibfnamefont {R.~A.}\ \bibnamefont {Simha}},\ }\href {\doibase
  10.1103/RevModPhys.85.1143} {\bibfield  {journal} {\bibinfo  {journal}
  {Reviews of Modern Physics}\ }\textbf {\bibinfo {volume} {85}},\ \bibinfo
  {pages} {1143} (\bibinfo {year} {2013})},\ \Eprint
  {http://arxiv.org/abs/1207.2929} {arXiv:1207.2929} \BibitemShut {NoStop}%
\bibitem [{\citenamefont {Prost}\ \emph {et~al.}(2015)\citenamefont {Prost},
  \citenamefont {J{\"{u}}licher},\ and\ \citenamefont {Joanny}}]{Prost2015}%
  \BibitemOpen
  \bibfield  {author} {\bibinfo {author} {\bibfnamefont {J.}~\bibnamefont
  {Prost}}, \bibinfo {author} {\bibfnamefont {F.}~\bibnamefont
  {J{\"{u}}licher}}, \ and\ \bibinfo {author} {\bibfnamefont {J.~F.}\
  \bibnamefont {Joanny}},\ }\href {\doibase 10.1038/nphys3224} {\bibfield
  {journal} {\bibinfo  {journal} {Nature Physics}\ }\textbf {\bibinfo {volume}
  {11}},\ \bibinfo {pages} {111} (\bibinfo {year} {2015})}\BibitemShut
  {NoStop}%
\bibitem [{\citenamefont {Solon}\ \emph {et~al.}(2015)\citenamefont {Solon},
  \citenamefont {Cates},\ and\ \citenamefont {Tailleur}}]{Solon2015}%
  \BibitemOpen
  \bibfield  {author} {\bibinfo {author} {\bibfnamefont {A.~P.}\ \bibnamefont
  {Solon}}, \bibinfo {author} {\bibfnamefont {M.~E.}\ \bibnamefont {Cates}}, \
  and\ \bibinfo {author} {\bibfnamefont {J.}~\bibnamefont {Tailleur}},\ }\href
  {\doibase 10.1140/epjst/e2015-02457-0} {\bibfield  {journal} {\bibinfo
  {journal} {European Physical Journal: Special Topics}\ }\textbf {\bibinfo
  {volume} {224}},\ \bibinfo {pages} {1231} (\bibinfo {year} {2015})},\ \Eprint
  {http://arxiv.org/abs/1504.07391} {arXiv:1504.07391} \BibitemShut {NoStop}%
\bibitem [{\citenamefont {Dupin}\ and\ \citenamefont
  {Etienne-Manneville}(2011)}]{Dupin2011}%
  \BibitemOpen
  \bibfield  {author} {\bibinfo {author} {\bibfnamefont {I.}~\bibnamefont
  {Dupin}}\ and\ \bibinfo {author} {\bibfnamefont {S.}~\bibnamefont
  {Etienne-Manneville}},\ }\href {\doibase 10.1016/j.biocel.2011.09.004}
  {\bibfield  {journal} {\bibinfo  {journal} {International Journal of
  Biochemistry and Cell Biology}\ }\textbf {\bibinfo {volume} {43}},\ \bibinfo
  {pages} {1698} (\bibinfo {year} {2011})},\ \Eprint
  {http://arxiv.org/abs/NIHMS150003} {arXiv:NIHMS150003} \BibitemShut {NoStop}%
\bibitem [{\citenamefont {Gundersen}\ and\ \citenamefont
  {Worman}(2013)}]{Gundersen2013}%
  \BibitemOpen
  \bibfield  {author} {\bibinfo {author} {\bibfnamefont {G.~G.}\ \bibnamefont
  {Gundersen}}\ and\ \bibinfo {author} {\bibfnamefont {H.~J.}\ \bibnamefont
  {Worman}},\ }\href {\doibase 10.1016/j.cell.2013.02.031} {\bibfield
  {journal} {\bibinfo  {journal} {Cell}\ }\textbf {\bibinfo {volume} {152}},\
  \bibinfo {pages} {1376} (\bibinfo {year} {2013})},\ \Eprint
  {http://arxiv.org/abs/NIHMS150003} {arXiv:NIHMS150003} \BibitemShut {NoStop}%
\bibitem [{\citenamefont {Morris}(2003)}]{Morris2002}%
  \BibitemOpen
  \bibfield  {author} {\bibinfo {author} {\bibfnamefont {N.~R.}\ \bibnamefont
  {Morris}},\ }\href {\doibase 10.1016/S0955-0674(02)00004-2} {\bibfield
  {journal} {\bibinfo  {journal} {Current Opinion in Cell Biology}\ }\textbf
  {\bibinfo {volume} {15}},\ \bibinfo {pages} {54} (\bibinfo {year}
  {2003})}\BibitemShut {NoStop}%
\bibitem [{\citenamefont {Makhija}\ \emph {et~al.}(2016)\citenamefont
  {Makhija}, \citenamefont {Jokhun},\ and\ \citenamefont
  {Shivashankar}}]{Makhija2016}%
  \BibitemOpen
  \bibfield  {author} {\bibinfo {author} {\bibfnamefont {E.}~\bibnamefont
  {Makhija}}, \bibinfo {author} {\bibfnamefont {D.~S.}\ \bibnamefont {Jokhun}},
  \ and\ \bibinfo {author} {\bibfnamefont {G.~V.}\ \bibnamefont
  {Shivashankar}},\ }\href {\doibase 10.1073/pnas.1513189113} {\bibfield
  {journal} {\bibinfo  {journal} {Proceedings of the National Academy of
  Sciences}\ }\textbf {\bibinfo {volume} {113}},\ \bibinfo {pages} {E32}
  (\bibinfo {year} {2016})}\BibitemShut {NoStop}%
\bibitem [{\citenamefont {Talwar}\ \emph {et~al.}(2013)\citenamefont {Talwar},
  \citenamefont {Kumar}, \citenamefont {Rao}, \citenamefont {Menon},\ and\
  \citenamefont {Shivashankar}}]{Talwar2013}%
  \BibitemOpen
  \bibfield  {author} {\bibinfo {author} {\bibfnamefont {S.}~\bibnamefont
  {Talwar}}, \bibinfo {author} {\bibfnamefont {A.}~\bibnamefont {Kumar}},
  \bibinfo {author} {\bibfnamefont {M.}~\bibnamefont {Rao}}, \bibinfo {author}
  {\bibfnamefont {G.~I.}\ \bibnamefont {Menon}}, \ and\ \bibinfo {author}
  {\bibfnamefont {G.~V.}\ \bibnamefont {Shivashankar}},\ }\href {\doibase
  10.1016/j.bpj.2012.12.033} {\bibfield  {journal} {\bibinfo  {journal}
  {Biophysical Journal}\ }\textbf {\bibinfo {volume} {104}},\ \bibinfo {pages}
  {553} (\bibinfo {year} {2013})}\BibitemShut {NoStop}%
\bibitem [{\citenamefont {Weihs}\ \emph {et~al.}(2006)\citenamefont {Weihs},
  \citenamefont {Mason},\ and\ \citenamefont {Teitell}}]{Weihs2006}%
  \BibitemOpen
  \bibfield  {author} {\bibinfo {author} {\bibfnamefont {D.}~\bibnamefont
  {Weihs}}, \bibinfo {author} {\bibfnamefont {T.~G.}\ \bibnamefont {Mason}}, \
  and\ \bibinfo {author} {\bibfnamefont {M.~A.}\ \bibnamefont {Teitell}},\
  }\href {\doibase 10.1529/biophysj.106.081109} {\bibfield  {journal} {\bibinfo
   {journal} {Biophysical Journal}\ }\textbf {\bibinfo {volume} {91}},\
  \bibinfo {pages} {4296} (\bibinfo {year} {2006})}\BibitemShut {NoStop}%
\bibitem [{\citenamefont {Hameed}\ \emph {et~al.}(2012)\citenamefont {Hameed},
  \citenamefont {Rao},\ and\ \citenamefont {Shivashankar}}]{Hameed2012}%
  \BibitemOpen
  \bibfield  {author} {\bibinfo {author} {\bibfnamefont {F.~M.}\ \bibnamefont
  {Hameed}}, \bibinfo {author} {\bibfnamefont {M.}~\bibnamefont {Rao}}, \ and\
  \bibinfo {author} {\bibfnamefont {G.~V.}\ \bibnamefont {Shivashankar}},\
  }\href {\doibase 10.1371/journal.pone.0045843} {\bibfield  {journal}
  {\bibinfo  {journal} {PLoS ONE}\ }\textbf {\bibinfo {volume} {7}},\ \bibinfo
  {pages} {e45843} (\bibinfo {year} {2012})}\BibitemShut {NoStop}%
\bibitem [{\citenamefont {Lau}\ \emph {et~al.}(2003)\citenamefont {Lau},
  \citenamefont {Hoffman}, \citenamefont {Davies}, \citenamefont {Crocker},\
  and\ \citenamefont {Lubensky}}]{Lau2003}%
  \BibitemOpen
  \bibfield  {author} {\bibinfo {author} {\bibfnamefont {A.~W.}\ \bibnamefont
  {Lau}}, \bibinfo {author} {\bibfnamefont {B.~D.}\ \bibnamefont {Hoffman}},
  \bibinfo {author} {\bibfnamefont {A.}~\bibnamefont {Davies}}, \bibinfo
  {author} {\bibfnamefont {J.~C.}\ \bibnamefont {Crocker}}, \ and\ \bibinfo
  {author} {\bibfnamefont {T.~C.}\ \bibnamefont {Lubensky}},\ }\href {\doibase
  10.1103/PhysRevLett.91.198101} {\bibfield  {journal} {\bibinfo  {journal}
  {Physical Review Letters}\ }\textbf {\bibinfo {volume} {91}},\ \bibinfo
  {pages} {7} (\bibinfo {year} {2003})},\ \Eprint
  {http://arxiv.org/abs/0309510} {arXiv:0309510 [cond-mat]} \BibitemShut
  {NoStop}%
\bibitem [{\citenamefont {Fodor}\ \emph {et~al.}(2015)\citenamefont {Fodor},
  \citenamefont {Guo}, \citenamefont {Gov}, \citenamefont {Visco},
  \citenamefont {Weitz},\ and\ \citenamefont {{Van Wijland}}}]{Fodor2015}%
  \BibitemOpen
  \bibfield  {author} {\bibinfo {author} {\bibnamefont {Fodor}}, \bibinfo
  {author} {\bibfnamefont {M.}~\bibnamefont {Guo}}, \bibinfo {author}
  {\bibfnamefont {N.~S.}\ \bibnamefont {Gov}}, \bibinfo {author} {\bibfnamefont
  {P.}~\bibnamefont {Visco}}, \bibinfo {author} {\bibfnamefont {D.~A.}\
  \bibnamefont {Weitz}}, \ and\ \bibinfo {author} {\bibfnamefont
  {F.}~\bibnamefont {{Van Wijland}}},\ }\href {\doibase
  10.1209/0295-5075/110/48005} {\bibfield  {journal} {\bibinfo  {journal}
  {Epl}\ }\textbf {\bibinfo {volume} {110}},\ \bibinfo {pages} {48005}
  (\bibinfo {year} {2015})},\ \Eprint {http://arxiv.org/abs/arXiv:1505.06489v1}
  {arXiv:arXiv:1505.06489v1} \BibitemShut {NoStop}%
\bibitem [{\citenamefont {Mazumdar}\ and\ \citenamefont
  {Mazumdar}(2002)}]{Mazumdar2002}%
  \BibitemOpen
  \bibfield  {author} {\bibinfo {author} {\bibfnamefont {A.}~\bibnamefont
  {Mazumdar}}\ and\ \bibinfo {author} {\bibfnamefont {M.}~\bibnamefont
  {Mazumdar}},\ }\href {\doibase 10.1002/bies.10184} {\bibfield  {journal}
  {\bibinfo  {journal} {BioEssays}\ }\textbf {\bibinfo {volume} {24}},\
  \bibinfo {pages} {1012} (\bibinfo {year} {2002})}\BibitemShut {NoStop}%
\bibitem [{\citenamefont {Cremer}\ and\ \citenamefont
  {Cremer}(2001)}]{Cremer2001}%
  \BibitemOpen
  \bibfield  {author} {\bibinfo {author} {\bibfnamefont {T.}~\bibnamefont
  {Cremer}}\ and\ \bibinfo {author} {\bibfnamefont {C.}~\bibnamefont
  {Cremer}},\ }\href {\doibase 10.1038/35066075} {\bibfield  {journal}
  {\bibinfo  {journal} {Nature Reviews Genetics}\ }\textbf {\bibinfo {volume}
  {2}},\ \bibinfo {pages} {292} (\bibinfo {year} {2001})}\BibitemShut {NoStop}%
\bibitem [{\citenamefont {Bruinsma}\ \emph {et~al.}(2014)\citenamefont
  {Bruinsma}, \citenamefont {Grosberg}, \citenamefont {Rabin},\ and\
  \citenamefont {Zidovska}}]{Bruinsma2014}%
  \BibitemOpen
  \bibfield  {author} {\bibinfo {author} {\bibfnamefont {R.}~\bibnamefont
  {Bruinsma}}, \bibinfo {author} {\bibfnamefont {A.~Y.}\ \bibnamefont
  {Grosberg}}, \bibinfo {author} {\bibfnamefont {Y.}~\bibnamefont {Rabin}}, \
  and\ \bibinfo {author} {\bibfnamefont {A.}~\bibnamefont {Zidovska}},\ }\href
  {\doibase 10.1016/j.bpj.2014.03.038} {\bibfield  {journal} {\bibinfo
  {journal} {Biophysical Journal}\ }\textbf {\bibinfo {volume} {106}},\
  \bibinfo {pages} {1871} (\bibinfo {year} {2014})}\BibitemShut {NoStop}%
\bibitem [{\citenamefont {Weber}\ \emph {et~al.}(2012)\citenamefont {Weber},
  \citenamefont {Spakowitz},\ and\ \citenamefont {Theriot}}]{Weber2012}%
  \BibitemOpen
  \bibfield  {author} {\bibinfo {author} {\bibfnamefont {S.~C.}\ \bibnamefont
  {Weber}}, \bibinfo {author} {\bibfnamefont {A.~J.}\ \bibnamefont
  {Spakowitz}}, \ and\ \bibinfo {author} {\bibfnamefont {J.~A.}\ \bibnamefont
  {Theriot}},\ }\href {\doibase 10.1073/pnas.1119505109} {\bibfield  {journal}
  {\bibinfo  {journal} {Proceedings of the National Academy of Sciences}\
  }\textbf {\bibinfo {volume} {109}},\ \bibinfo {pages} {7338} (\bibinfo {year}
  {2012})}\BibitemShut {NoStop}%
\bibitem [{\citenamefont {Ganai}\ \emph {et~al.}(2014)\citenamefont {Ganai},
  \citenamefont {Sengupta},\ and\ \citenamefont {Menon}}]{Ganai2014}%
  \BibitemOpen
  \bibfield  {author} {\bibinfo {author} {\bibfnamefont {N.}~\bibnamefont
  {Ganai}}, \bibinfo {author} {\bibfnamefont {S.}~\bibnamefont {Sengupta}}, \
  and\ \bibinfo {author} {\bibfnamefont {G.~I.}\ \bibnamefont {Menon}},\ }\href
  {\doibase 10.1093/nar/gkt1417} {\bibfield  {journal} {\bibinfo  {journal}
  {Nucleic Acids Research}\ }\textbf {\bibinfo {volume} {42}},\ \bibinfo
  {pages} {4145} (\bibinfo {year} {2014})}\BibitemShut {NoStop}%
\bibitem [{\citenamefont {Barton}\ \emph {et~al.}(2017)\citenamefont {Barton},
  \citenamefont {Henkes}, \citenamefont {Weijer},\ and\ \citenamefont
  {Sknepnek}}]{Barton2017}%
  \BibitemOpen
  \bibfield  {author} {\bibinfo {author} {\bibfnamefont {D.~L.}\ \bibnamefont
  {Barton}}, \bibinfo {author} {\bibfnamefont {S.}~\bibnamefont {Henkes}},
  \bibinfo {author} {\bibfnamefont {C.~J.}\ \bibnamefont {Weijer}}, \ and\
  \bibinfo {author} {\bibfnamefont {R.}~\bibnamefont {Sknepnek}},\ }\href
  {\doibase 10.1371/journal.pcbi.1005569} {\bibfield  {journal} {\bibinfo
  {journal} {PLoS Computational Biology}\ }\textbf {\bibinfo {volume} {13}},\
  \bibinfo {pages} {e1005569} (\bibinfo {year} {2017})},\ \Eprint
  {http://arxiv.org/abs/1612.05960} {arXiv:1612.05960} \BibitemShut {NoStop}%
\bibitem [{\citenamefont {Curran}\ \emph {et~al.}(2017)\citenamefont {Curran},
  \citenamefont {Strandkvist}, \citenamefont {Bathmann}, \citenamefont
  {de~Gennes}, \citenamefont {Kabla}, \citenamefont {Salbreux},\ and\
  \citenamefont {Baum}}]{Curran2017}%
  \BibitemOpen
  \bibfield  {author} {\bibinfo {author} {\bibfnamefont {S.}~\bibnamefont
  {Curran}}, \bibinfo {author} {\bibfnamefont {C.}~\bibnamefont {Strandkvist}},
  \bibinfo {author} {\bibfnamefont {J.}~\bibnamefont {Bathmann}}, \bibinfo
  {author} {\bibfnamefont {M.}~\bibnamefont {de~Gennes}}, \bibinfo {author}
  {\bibfnamefont {A.}~\bibnamefont {Kabla}}, \bibinfo {author} {\bibfnamefont
  {G.}~\bibnamefont {Salbreux}}, \ and\ \bibinfo {author} {\bibfnamefont
  {B.}~\bibnamefont {Baum}},\ }\href {\doibase 10.1016/j.devcel.2017.09.018}
  {\bibfield  {journal} {\bibinfo  {journal} {Developmental Cell}\ }\textbf
  {\bibinfo {volume} {43}},\ \bibinfo {pages} {480} (\bibinfo {year}
  {2017})}\BibitemShut {NoStop}%
\bibitem [{\citenamefont {Li}\ \emph {et~al.}(2014)\citenamefont {Li},
  \citenamefont {Kumar}, \citenamefont {Makhija},\ and\ \citenamefont
  {Shivashankar}}]{Li2014}%
  \BibitemOpen
  \bibfield  {author} {\bibinfo {author} {\bibfnamefont {Q.}~\bibnamefont
  {Li}}, \bibinfo {author} {\bibfnamefont {A.}~\bibnamefont {Kumar}}, \bibinfo
  {author} {\bibfnamefont {E.}~\bibnamefont {Makhija}}, \ and\ \bibinfo
  {author} {\bibfnamefont {G.~V.}\ \bibnamefont {Shivashankar}},\ }\href
  {\doibase 10.1016/j.biomaterials.2013.10.037} {\bibfield  {journal} {\bibinfo
   {journal} {Biomaterials}\ }\textbf {\bibinfo {volume} {35}},\ \bibinfo
  {pages} {961} (\bibinfo {year} {2014})}\BibitemShut {NoStop}%
\bibitem [{\citenamefont {Rupprecht}\ \emph {et~al.}(2018)\citenamefont
  {Rupprecht}, \citenamefont {Singh~Vishen}, \citenamefont {Shivashankar},
  \citenamefont {Rao},\ and\ \citenamefont {Prost}}]{Rupprecht2017}%
  \BibitemOpen
  \bibfield  {author} {\bibinfo {author} {\bibfnamefont {J.-F.}\ \bibnamefont
  {Rupprecht}}, \bibinfo {author} {\bibfnamefont {A.}~\bibnamefont
  {Singh~Vishen}}, \bibinfo {author} {\bibfnamefont {G.~V.}\ \bibnamefont
  {Shivashankar}}, \bibinfo {author} {\bibfnamefont {M.}~\bibnamefont {Rao}}, \
  and\ \bibinfo {author} {\bibfnamefont {J.}~\bibnamefont {Prost}},\ }\href
  {\doibase 10.1103/PhysRevLett.120.098001} {\bibfield  {journal} {\bibinfo
  {journal} {Phys. Rev. Lett.}\ }\textbf {\bibinfo {volume} {120}},\ \bibinfo
  {pages} {098001} (\bibinfo {year} {2018})}\BibitemShut {NoStop}%
\bibitem [{\citenamefont {Bartolo}\ \emph {et~al.}(2003)\citenamefont
  {Bartolo}, \citenamefont {Ajdari},\ and\ \citenamefont
  {Fournier}}]{Bartolo2003}%
  \BibitemOpen
  \bibfield  {author} {\bibinfo {author} {\bibfnamefont {D.}~\bibnamefont
  {Bartolo}}, \bibinfo {author} {\bibfnamefont {A.}~\bibnamefont {Ajdari}}, \
  and\ \bibinfo {author} {\bibfnamefont {J.~B.}\ \bibnamefont {Fournier}},\
  }\href {\doibase 10.1103/PhysRevE.67.061112} {\bibfield  {journal} {\bibinfo
  {journal} {Physical Review E - Statistical Physics, Plasmas, Fluids, and
  Related Interdisciplinary Topics}\ }\textbf {\bibinfo {volume} {67}},\
  \bibinfo {pages} {9} (\bibinfo {year} {2003})},\ \Eprint
  {http://arxiv.org/abs/0304356} {arXiv:0304356 [cond-mat]} \BibitemShut
  {NoStop}%
\bibitem [{\citenamefont {Ray}\ \emph {et~al.}(2014)\citenamefont {Ray},
  \citenamefont {Reichhardt},\ and\ \citenamefont {Reichhardt}}]{Ray2014}%
  \BibitemOpen
  \bibfield  {author} {\bibinfo {author} {\bibfnamefont {D.}~\bibnamefont
  {Ray}}, \bibinfo {author} {\bibfnamefont {C.}~\bibnamefont {Reichhardt}}, \
  and\ \bibinfo {author} {\bibfnamefont {C.~J.~O.}\ \bibnamefont
  {Reichhardt}},\ }\href {\doibase 10.1103/PhysRevE.90.013019} {\bibfield
  {journal} {\bibinfo  {journal} {Physical Review E - Statistical, Nonlinear,
  and Soft Matter Physics}\ }\textbf {\bibinfo {volume} {90}},\ \bibinfo
  {pages} {1} (\bibinfo {year} {2014})},\ \Eprint
  {http://arxiv.org/abs/1402.6372} {arXiv:1402.6372} \BibitemShut {NoStop}%
\bibitem [{\citenamefont {Parra-Rojas}\ and\ \citenamefont
  {Soto}(2014)}]{Parra-Rojas2014}%
  \BibitemOpen
  \bibfield  {author} {\bibinfo {author} {\bibfnamefont {C.}~\bibnamefont
  {Parra-Rojas}}\ and\ \bibinfo {author} {\bibfnamefont {R.}~\bibnamefont
  {Soto}},\ }\href {\doibase 10.1103/PhysRevE.90.013024} {\bibfield  {journal}
  {\bibinfo  {journal} {Physical Review E - Statistical, Nonlinear, and Soft
  Matter Physics}\ }\textbf {\bibinfo {volume} {90}},\ \bibinfo {pages} {1}
  (\bibinfo {year} {2014})},\ \Eprint {http://arxiv.org/abs/1404.4857}
  {arXiv:1404.4857} \BibitemShut {NoStop}%
\bibitem [{\citenamefont {Kruse}\ \emph {et~al.}(2005)\citenamefont {Kruse},
  \citenamefont {Joanny}, \citenamefont {J{\"{u}}licher}, \citenamefont
  {Prost},\ and\ \citenamefont {Sekimoto}}]{Kruse2005}%
  \BibitemOpen
  \bibfield  {author} {\bibinfo {author} {\bibfnamefont {K.}~\bibnamefont
  {Kruse}}, \bibinfo {author} {\bibfnamefont {J.~F.}\ \bibnamefont {Joanny}},
  \bibinfo {author} {\bibfnamefont {F.}~\bibnamefont {J{\"{u}}licher}},
  \bibinfo {author} {\bibfnamefont {J.}~\bibnamefont {Prost}}, \ and\ \bibinfo
  {author} {\bibfnamefont {K.}~\bibnamefont {Sekimoto}},\ }\href {\doibase
  10.1140/epje/e2005-00002-5} {\bibfield  {journal} {\bibinfo  {journal}
  {European Physical Journal E}\ }\textbf {\bibinfo {volume} {16}},\ \bibinfo
  {pages} {5} (\bibinfo {year} {2005})},\ \Eprint
  {http://arxiv.org/abs/0406058} {arXiv:0406058 [physics]} \BibitemShut
  {NoStop}%
\bibitem [{\citenamefont {Basu}\ \emph {et~al.}(2008)\citenamefont {Basu},
  \citenamefont {Joanny}, \citenamefont {J{\"{u}}licher},\ and\ \citenamefont
  {Prost}}]{Basu2008}%
  \BibitemOpen
  \bibfield  {author} {\bibinfo {author} {\bibfnamefont {A.}~\bibnamefont
  {Basu}}, \bibinfo {author} {\bibfnamefont {J.~F.}\ \bibnamefont {Joanny}},
  \bibinfo {author} {\bibfnamefont {F.}~\bibnamefont {J{\"{u}}licher}}, \ and\
  \bibinfo {author} {\bibfnamefont {J.}~\bibnamefont {Prost}},\ }\href
  {\doibase 10.1140/epje/i2008-10364-9} {\bibfield  {journal} {\bibinfo
  {journal} {European Physical Journal E}\ }\textbf {\bibinfo {volume} {27}},\
  \bibinfo {pages} {149} (\bibinfo {year} {2008})}\BibitemShut {NoStop}%
\bibitem [{\citenamefont {Saha}\ \emph {et~al.}(2016)\citenamefont {Saha},
  \citenamefont {Nishikawa}, \citenamefont {Behrndt}, \citenamefont
  {Heisenberg}, \citenamefont {J{\"{u}}licher},\ and\ \citenamefont
  {Grill}}]{Saha2016}%
  \BibitemOpen
  \bibfield  {author} {\bibinfo {author} {\bibfnamefont {A.}~\bibnamefont
  {Saha}}, \bibinfo {author} {\bibfnamefont {M.}~\bibnamefont {Nishikawa}},
  \bibinfo {author} {\bibfnamefont {M.}~\bibnamefont {Behrndt}}, \bibinfo
  {author} {\bibfnamefont {C.~P.}\ \bibnamefont {Heisenberg}}, \bibinfo
  {author} {\bibfnamefont {F.}~\bibnamefont {J{\"{u}}licher}}, \ and\ \bibinfo
  {author} {\bibfnamefont {S.~W.}\ \bibnamefont {Grill}},\ }\href {\doibase
  10.1016/j.bpj.2016.02.013} {\bibfield  {journal} {\bibinfo  {journal}
  {Biophysical Journal}\ }\textbf {\bibinfo {volume} {110}},\ \bibinfo {pages}
  {1421} (\bibinfo {year} {2016})},\ \Eprint {http://arxiv.org/abs/1507.00511}
  {arXiv:1507.00511} \BibitemShut {NoStop}%
\bibitem [{\citenamefont {Gardiner}(2009)}]{Gardiner}%
  \BibitemOpen
  \bibfield  {author} {\bibinfo {author} {\bibfnamefont {C.}~\bibnamefont
  {Gardiner}},\ }\href@noop {} {\emph {\bibinfo {title} {Springer}}},\ \bibinfo
  {edition} {4th}\ ed.\ (\bibinfo  {publisher} {Springer},\ \bibinfo {year}
  {2009})\ p.\ \bibinfo {pages} {447}\BibitemShut {NoStop}%
\bibitem [{\citenamefont {van Kampen}(1981)}]{Kampen1981}%
  \BibitemOpen
  \bibfield  {author} {\bibinfo {author} {\bibfnamefont {N.~G.}\ \bibnamefont
  {van Kampen}},\ }\href {\doibase 10.1007/BF01007642} {\bibfield  {journal}
  {\bibinfo  {journal} {Journal of Statistical Physics}\ }\textbf {\bibinfo
  {volume} {24}},\ \bibinfo {pages} {175} (\bibinfo {year} {1981})}\BibitemShut
  {NoStop}%
\bibitem [{\citenamefont {Kupferman}\ \emph {et~al.}(2004)\citenamefont
  {Kupferman}, \citenamefont {Pavliotis},\ and\ \citenamefont
  {Stuart}}]{Kupferman2004}%
  \BibitemOpen
  \bibfield  {author} {\bibinfo {author} {\bibfnamefont {R.}~\bibnamefont
  {Kupferman}}, \bibinfo {author} {\bibfnamefont {G.~A.}\ \bibnamefont
  {Pavliotis}}, \ and\ \bibinfo {author} {\bibfnamefont {A.~M.}\ \bibnamefont
  {Stuart}},\ }\href {\doibase 10.1103/PhysRevE.70.036120} {\bibfield
  {journal} {\bibinfo  {journal} {Physical Review E - Statistical Physics,
  Plasmas, Fluids, and Related Interdisciplinary Topics}\ }\textbf {\bibinfo
  {volume} {70}},\ \bibinfo {pages} {9} (\bibinfo {year} {2004})}\BibitemShut
  {NoStop}%
\bibitem [{\citenamefont {Sancho}\ \emph {et~al.}(1982)\citenamefont {Sancho},
  \citenamefont {Miguel},\ and\ \citenamefont {D{\"u}rr}}]{Sancho1982}%
  \BibitemOpen
  \bibfield  {author} {\bibinfo {author} {\bibfnamefont {J.~M.}\ \bibnamefont
  {Sancho}}, \bibinfo {author} {\bibfnamefont {M.~S.}\ \bibnamefont {Miguel}},
  \ and\ \bibinfo {author} {\bibfnamefont {D.}~\bibnamefont {D{\"u}rr}},\
  }\href {\doibase 10.1007/BF01012607} {\bibfield  {journal} {\bibinfo
  {journal} {Journal of Statistical Physics}\ }\textbf {\bibinfo {volume}
  {28}},\ \bibinfo {pages} {291} (\bibinfo {year} {1982})}\BibitemShut
  {NoStop}%
\bibitem [{\citenamefont {Lau}\ and\ \citenamefont {Lubensky}(2007)}]{Lau2007}%
  \BibitemOpen
  \bibfield  {author} {\bibinfo {author} {\bibfnamefont {A.~W.~C.}\
  \bibnamefont {Lau}}\ and\ \bibinfo {author} {\bibfnamefont {T.~C.}\
  \bibnamefont {Lubensky}},\ }\href {\doibase 10.1103/PhysRevE.76.011123}
  {\bibfield  {journal} {\bibinfo  {journal} {Physical Review E}\ }\textbf
  {\bibinfo {volume} {76}},\ \bibinfo {pages} {011123} (\bibinfo {year}
  {2007})},\ \Eprint {http://arxiv.org/abs/0707.2234} {arXiv:0707.2234}
  \BibitemShut {NoStop}%
\bibitem [{\citenamefont {H{\"{a}}nggi}\ and\ \citenamefont
  {Thomas}(1982)}]{Hanggi1982}%
  \BibitemOpen
  \bibfield  {author} {\bibinfo {author} {\bibfnamefont {P.}~\bibnamefont
  {H{\"{a}}nggi}}\ and\ \bibinfo {author} {\bibfnamefont {H.}~\bibnamefont
  {Thomas}},\ }\href {\doibase 10.1016/0370-1573(82)90045-X} {\bibfield
  {journal} {\bibinfo  {journal} {Physics Reports}\ }\textbf {\bibinfo {volume}
  {88}},\ \bibinfo {pages} {207} (\bibinfo {year} {1982})}\BibitemShut
  {NoStop}%
\bibitem [{\citenamefont {Klimontovich}(1990)}]{Klimontovich1990}%
  \BibitemOpen
  \bibfield  {author} {\bibinfo {author} {\bibfnamefont {Y.~L.}\ \bibnamefont
  {Klimontovich}},\ }\href {\doibase 10.1016/0378-4371(90)90142-F} {\bibfield
  {journal} {\bibinfo  {journal} {Physica A: Statistical Mechanics and its
  Applications}\ }\textbf {\bibinfo {volume} {163}},\ \bibinfo {pages} {515}
  (\bibinfo {year} {1990})}\BibitemShut {NoStop}%
\bibitem [{\citenamefont {Stratonovich}(1966)}]{Stratonovich}%
  \BibitemOpen
  \bibfield  {author} {\bibinfo {author} {\bibfnamefont {R.~L.}\ \bibnamefont
  {Stratonovich}},\ }\href {http://epubs.siam.org/doi/pdf/10.1137/0304028}
  {\bibfield  {journal} {\bibinfo  {journal} {J. SIAM Control}\ }\textbf
  {\bibinfo {volume} {4}},\ \bibinfo {pages} {362} (\bibinfo {year}
  {1966})}\BibitemShut {NoStop}%
\bibitem [{\citenamefont {Versaevel}\ \emph {et~al.}(2012)\citenamefont
  {Versaevel}, \citenamefont {Grevesse},\ and\ \citenamefont
  {Gabriele}}]{Versaevel2012}%
  \BibitemOpen
  \bibfield  {author} {\bibinfo {author} {\bibfnamefont {M.}~\bibnamefont
  {Versaevel}}, \bibinfo {author} {\bibfnamefont {T.}~\bibnamefont {Grevesse}},
  \ and\ \bibinfo {author} {\bibfnamefont {S.}~\bibnamefont {Gabriele}},\
  }\href {\doibase 10.1038/ncomms1668} {\bibfield  {journal} {\bibinfo
  {journal} {Nature Communications}\ }\textbf {\bibinfo {volume} {3}},\
  \bibinfo {pages} {671} (\bibinfo {year} {2012})}\BibitemShut {NoStop}%
\bibitem [{\citenamefont {{De Gennes}}(1985)}]{DeGennes1985}%
  \BibitemOpen
  \bibfield  {author} {\bibinfo {author} {\bibfnamefont {P.~G.}\ \bibnamefont
  {{De Gennes}}},\ }\href {\doibase 10.1103/RevModPhys.57.827} {\bibfield
  {journal} {\bibinfo  {journal} {Reviews of Modern Physics}\ }\textbf
  {\bibinfo {volume} {57}},\ \bibinfo {pages} {827} (\bibinfo {year}
  {1985})}\BibitemShut {NoStop}%
\bibitem [{\citenamefont {Arfken}\ \emph {et~al.}(2012)\citenamefont {Arfken},
  \citenamefont {Weber},\ and\ \citenamefont {Harris}}]{Arfken}%
  \BibitemOpen
  \bibfield  {author} {\bibinfo {author} {\bibfnamefont {G.}~\bibnamefont
  {Arfken}}, \bibinfo {author} {\bibfnamefont {H.}~\bibnamefont {Weber}}, \
  and\ \bibinfo {author} {\bibfnamefont {F.~E.}\ \bibnamefont {Harris}},\
  }\href
  {http://books.google.com.hk/books/about/Solitons.html?id=WEYsAAAAYAAJ{\&}pgis=1}
  {\emph {\bibinfo {title} {{Mathematical Methods for Physicists}}}},\ \bibinfo
  {edition} {7th}\ ed.\ (\bibinfo  {publisher} {Academic Press},\ \bibinfo
  {address} {Orlando FL},\ \bibinfo {year} {2012})\BibitemShut {NoStop}%
\bibitem [{\citenamefont {Ramdas}\ and\ \citenamefont
  {Shivashankar}(2015)}]{Ramdas2015}%
  \BibitemOpen
  \bibfield  {author} {\bibinfo {author} {\bibfnamefont {N.~M.}\ \bibnamefont
  {Ramdas}}\ and\ \bibinfo {author} {\bibfnamefont {G.~V.}\ \bibnamefont
  {Shivashankar}},\ }\href {\doibase 10.1016/j.jmb.2014.09.008} {\bibfield
  {journal} {\bibinfo  {journal} {Journal of Molecular Biology}\ }\textbf
  {\bibinfo {volume} {427}},\ \bibinfo {pages} {695} (\bibinfo {year}
  {2015})}\BibitemShut {NoStop}%
\bibitem [{\citenamefont {Aminov}\ \emph {et~al.}(2015)\citenamefont {Aminov},
  \citenamefont {Kafri},\ and\ \citenamefont {Kardar}}]{Aminov2015}%
  \BibitemOpen
  \bibfield  {author} {\bibinfo {author} {\bibfnamefont {A.}~\bibnamefont
  {Aminov}}, \bibinfo {author} {\bibfnamefont {Y.}~\bibnamefont {Kafri}}, \
  and\ \bibinfo {author} {\bibfnamefont {M.}~\bibnamefont {Kardar}},\ }\href
  {\doibase 10.1103/PhysRevLett.114.230602} {\bibfield  {journal} {\bibinfo
  {journal} {Physical Review Letters}\ }\textbf {\bibinfo {volume} {114}},\
  \bibinfo {pages} {230602} (\bibinfo {year} {2015})},\ \Eprint
  {http://arxiv.org/abs/1501.01006} {arXiv:1501.01006} \BibitemShut {NoStop}%
\bibitem [{\citenamefont {Brito}\ \emph {et~al.}(2007)\citenamefont {Brito},
  \citenamefont {{Marini Bettolo Marconi}},\ and\ \citenamefont
  {Soto}}]{Brito2007}%
  \BibitemOpen
  \bibfield  {author} {\bibinfo {author} {\bibfnamefont {R.}~\bibnamefont
  {Brito}}, \bibinfo {author} {\bibfnamefont {U.}~\bibnamefont {{Marini Bettolo
  Marconi}}}, \ and\ \bibinfo {author} {\bibfnamefont {R.}~\bibnamefont
  {Soto}},\ }\href {\doibase 10.1103/PhysRevE.76.011113} {\bibfield  {journal}
  {\bibinfo  {journal} {Physical Review E - Statistical, Nonlinear, and Soft
  Matter Physics}\ }\textbf {\bibinfo {volume} {76}},\ \bibinfo {pages} {1}
  (\bibinfo {year} {2007})},\ \Eprint {http://arxiv.org/abs/0611233}
  {arXiv:0611233 [cond-mat]} \BibitemShut {NoStop}%
\bibitem [{\citenamefont {Kirkpatrick}\ \emph {et~al.}(2013)\citenamefont
  {Kirkpatrick}, \citenamefont {{Ortiz De Z{\'{a}}rate}},\ and\ \citenamefont
  {Sengers}}]{Kirkpatrick2013}%
  \BibitemOpen
  \bibfield  {author} {\bibinfo {author} {\bibfnamefont {T.~R.}\ \bibnamefont
  {Kirkpatrick}}, \bibinfo {author} {\bibfnamefont {J.~M.}\ \bibnamefont
  {{Ortiz De Z{\'{a}}rate}}}, \ and\ \bibinfo {author} {\bibfnamefont {J.~V.}\
  \bibnamefont {Sengers}},\ }\href {\doibase 10.1103/PhysRevLett.110.235902}
  {\bibfield  {journal} {\bibinfo  {journal} {Physical Review Letters}\
  }\textbf {\bibinfo {volume} {110}},\ \bibinfo {pages} {1} (\bibinfo {year}
  {2013})},\ \Eprint {http://arxiv.org/abs/arXiv:1302.4704v1}
  {arXiv:arXiv:1302.4704v1} \BibitemShut {NoStop}%
\bibitem [{\citenamefont {Monahan}\ \emph {et~al.}(2016)\citenamefont
  {Monahan}, \citenamefont {Naji}, \citenamefont {Horgan}, \citenamefont {Lu},\
  and\ \citenamefont {Podgornik}}]{Monahan2015}%
  \BibitemOpen
  \bibfield  {author} {\bibinfo {author} {\bibfnamefont {C.}~\bibnamefont
  {Monahan}}, \bibinfo {author} {\bibfnamefont {A.}~\bibnamefont {Naji}},
  \bibinfo {author} {\bibfnamefont {R.}~\bibnamefont {Horgan}}, \bibinfo
  {author} {\bibfnamefont {B.-S.}\ \bibnamefont {Lu}}, \ and\ \bibinfo {author}
  {\bibfnamefont {R.}~\bibnamefont {Podgornik}},\ }\href {\doibase
  10.1039/C5SM02346G} {\bibfield  {journal} {\bibinfo  {journal} {Soft Matter}\
  }\textbf {\bibinfo {volume} {12}},\ \bibinfo {pages} {441} (\bibinfo {year}
  {2016})},\ \Eprint {http://arxiv.org/abs/arXiv:1310.1965v3}
  {arXiv:arXiv:1310.1965v3} \BibitemShut {NoStop}%
\bibitem [{\citenamefont {Wang}\ \emph {et~al.}(2017)\citenamefont {Wang},
  \citenamefont {Nagarajan}, \citenamefont {Uhler},\ and\ \citenamefont
  {Shivashankar}}]{Wang2017}%
  \BibitemOpen
  \bibfield  {author} {\bibinfo {author} {\bibfnamefont {Y.}~\bibnamefont
  {Wang}}, \bibinfo {author} {\bibfnamefont {M.}~\bibnamefont {Nagarajan}},
  \bibinfo {author} {\bibfnamefont {C.}~\bibnamefont {Uhler}}, \ and\ \bibinfo
  {author} {\bibfnamefont {G.~V.}\ \bibnamefont {Shivashankar}},\ }\href
  {\doibase 10.1091/mbc.E16-12-0825} {\bibfield  {journal} {\bibinfo  {journal}
  {Molecular Biology of the Cell}\ }\textbf {\bibinfo {volume} {28}},\ \bibinfo
  {pages} {1997} (\bibinfo {year} {2017})}\BibitemShut {NoStop}%
\bibitem [{\citenamefont {Uhler}\ and\ \citenamefont
  {Shivashankar}(2017)}]{Uhler2017}%
  \BibitemOpen
  \bibfield  {author} {\bibinfo {author} {\bibfnamefont {C.}~\bibnamefont
  {Uhler}}\ and\ \bibinfo {author} {\bibfnamefont {G.~V.}\ \bibnamefont
  {Shivashankar}},\ }\href {\doibase 10.1016/j.tcb.2017.06.005} {\bibfield
  {journal} {\bibinfo  {journal} {Trends in Cell Biology}\ }\textbf {\bibinfo
  {volume} {27}},\ \bibinfo {pages} {810} (\bibinfo {year} {2017})}\BibitemShut
  {NoStop}%
\bibitem [{\citenamefont {Cao}\ \emph {et~al.}(2015)\citenamefont {Cao},
  \citenamefont {Lin}, \citenamefont {Driscoll}, \citenamefont
  {Franco-Barraza}, \citenamefont {Cukierman}, \citenamefont {Mauck},\ and\
  \citenamefont {Shenoy}}]{Cao2015}%
  \BibitemOpen
  \bibfield  {author} {\bibinfo {author} {\bibfnamefont {X.}~\bibnamefont
  {Cao}}, \bibinfo {author} {\bibfnamefont {Y.}~\bibnamefont {Lin}}, \bibinfo
  {author} {\bibfnamefont {T.~P.}\ \bibnamefont {Driscoll}}, \bibinfo {author}
  {\bibfnamefont {J.}~\bibnamefont {Franco-Barraza}}, \bibinfo {author}
  {\bibfnamefont {E.}~\bibnamefont {Cukierman}}, \bibinfo {author}
  {\bibfnamefont {R.~L.}\ \bibnamefont {Mauck}}, \ and\ \bibinfo {author}
  {\bibfnamefont {V.~B.}\ \bibnamefont {Shenoy}},\ }\href {\doibase
  10.1016/j.bpj.2015.08.048} {\bibfield  {journal} {\bibinfo  {journal}
  {Biophysical Journal}\ }\textbf {\bibinfo {volume} {109}},\ \bibinfo {pages}
  {1807} (\bibinfo {year} {2015})}\BibitemShut {NoStop}%
\bibitem [{\citenamefont {Happel}\ and\ \citenamefont
  {Brenner}(1981)}]{Happel}%
  \BibitemOpen
  \bibfield  {author} {\bibinfo {author} {\bibfnamefont {J.}~\bibnamefont
  {Happel}}\ and\ \bibinfo {author} {\bibfnamefont {H.}~\bibnamefont
  {Brenner}},\ }\href {\doibase 10.1007/978-94-009-8352-6} {\emph {\bibinfo
  {title} {{Low Reynolds number hydrodynamics}}}},\ \bibinfo {series}
  {Mechanics of fluid and transport processes}, Vol.~\bibinfo {volume} {1}\
  (\bibinfo  {publisher} {Springer},\ \bibinfo {address} {Netherlands},\
  \bibinfo {year} {1981})\BibitemShut {NoStop}%
\bibitem [{\citenamefont {{M. Doi. and S. F. Edwards}}(1986)}]{Doi1988}%
  \BibitemOpen
  \bibfield  {author} {\bibinfo {author} {\bibnamefont {{M. Doi. and S. F.
  Edwards}}},\ }\href@noop {} {\emph {\bibinfo {title} {{The theory of polymer
  dynamics}}}},\ Vol.~\bibinfo {volume} {73}\ (\bibinfo  {publisher} {oxford
  university press},\ \bibinfo {year} {1986})\BibitemShut {NoStop}%
\bibitem [{\citenamefont {Fox}(1986)}]{Fox1986}%
  \BibitemOpen
  \bibfield  {author} {\bibinfo {author} {\bibfnamefont {R.~F.}\ \bibnamefont
  {Fox}},\ }\href {\doibase 10.1103/PhysRevA.33.467} {\bibfield  {journal}
  {\bibinfo  {journal} {Physical Review A}\ }\textbf {\bibinfo {volume} {33}},\
  \bibinfo {pages} {467} (\bibinfo {year} {1986})}\BibitemShut {NoStop}%
\bibitem [{\citenamefont {Baule}\ \emph {et~al.}(2008)\citenamefont {Baule},
  \citenamefont {{Vijay Kumar}},\ and\ \citenamefont {Ramaswamy}}]{Baule2009}%
  \BibitemOpen
  \bibfield  {author} {\bibinfo {author} {\bibfnamefont {A.}~\bibnamefont
  {Baule}}, \bibinfo {author} {\bibfnamefont {K.}~\bibnamefont {{Vijay
  Kumar}}}, \ and\ \bibinfo {author} {\bibfnamefont {S.}~\bibnamefont
  {Ramaswamy}},\ }\href {\doibase 10.1088/1742-5468/2008/11/P11008} {\bibfield
  {journal} {\bibinfo  {journal} {Journal of Statistical Mechanics: Theory and
  Experiment}\ }\textbf {\bibinfo {volume} {2008}},\ \bibinfo {pages} {11}
  (\bibinfo {year} {2008})},\ \Eprint {http://arxiv.org/abs/0903.1572}
  {arXiv:0903.1572} \BibitemShut {NoStop}%
\end{thebibliography}

%

\end{document}